\documentclass[%
preprint,
 amsmath,amssymb,
 aps,
]{revtex4-1}

\usepackage{graphicx} % needed for figures
\graphicspath{ {Figures/} }
\usepackage{epstopdf}
\usepackage[caption=false]{subfig}
\usepackage{amsmath}
\usepackage{amsfonts}
\usepackage{amssymb}
\usepackage{bm}
\usepackage{hyperref}
\usepackage{amsthm}
\usepackage{color}

\usepackage{mathrsfs}
\usepackage[vlined,ruled]{algorithm2e}
\usepackage{url}
\usepackage{color}
\usepackage{algorithmic}
\usepackage{bbm}
\usepackage{booktabs}
\usepackage{array}
\usepackage[table]{xcolor}
\usepackage{yfonts}

\newcommand{\mc}{\mathcal}

\newcommand{\real}{\mathbb{R}}

\newcommand\oprocendsymbol{\hbox{$\square$}}
\newcommand\oprocend{\relax\ifmmode\else\unskip\hfill\fi\oprocendsymbol}

\begin{document}
\bibliographystyle{naturemag}

\title{Conformational Control of Mechanical Networks}
\author{Jason Z. Kim}
\affiliation{Department of Bioengineering, University of Pennsylvania, Philadelphia, PA, 19104}
\author{Zhixin Lu}
\affiliation{Department of Bioengineering, University of Pennsylvania, Philadelphia, PA, 19104}
\author{Steven H. Strogatz}
\affiliation{Center for Applied Mathematics, Cornell University, Ithaca, NY, 14853, USA}
\author{Danielle S. Bassett}
\affiliation{Department of Bioengineering, University of Pennsylvania, Philadelphia, PA, 19104}
\affiliation{Department of Physics \& Astronomy, University of Pennsylvania, Philadelphia, PA, 19104}
\affiliation{Department of Electrical \& Systems Engineering, University of Pennsylvania, Philadelphia, PA, 19104}
\affiliation{To whom correspondence should be addressed: dsb@seas.upenn.edu}
\date{\today}
~\\
~\\
\begin{abstract}
Understanding conformational change is crucial for programming and controlling the function of many mechanical systems such as allosteric enzymes and tunable metamaterials. Of particular interest is the relationship between the network topology or geometry and the specific motions observed under controlling perturbations. We study this relationship in mechanical networks of 2-D and 3-D Maxwell frames composed of point masses connected by rigid rods rotating freely about the masses. We first develop simple principles that yield all bipartite network topologies and geometries that give rise to an arbitrarily specified instantaneous and finitely deformable motion in the masses as the sole non-rigid body zero mode. We then extend these principles to characterize networks that simultaneously yield multiple specified zero modes, and create large networks by coupling individual modules. These principles are then used to characterize and design networks with useful material (negative Poisson ratio) and mechanical (targeted allosteric response) functions.
\end{abstract}

\maketitle

\section{Introduction}

Many physical systems can be thought of as networks in which contacts, bonds, linkages, or hinges connect physical elements to one another. From the study of force chains in granular materials \cite{Papadopoulos2017network} to the study of fiber networks in polymer physics \cite{Picu2011mechanics}, it has become clear that both homogeneous and heterogeneous patterns of connectivity between physical elements can constrain the bulk properties of the material, including its response to stress and shear \cite{Vermeulen2017geometry}, its ability to transmit acoustic signals \cite{bassett2012influence}, and its capacity for thermal and electrical transport \cite{shi2014network}. These networks are also integral to the ever-evolving exploration of everyday machines \cite{norton2003design} in robotics \cite{detweiler2007robotics} and biology \cite{patek2007mantis}. Perhaps one of the simplest and most powerful conceptual advances in understanding such systems was the development of structural rigidity theory \cite{crapo1979structural}, built on a seminal early paper on constraint counting from J.C. Maxwell \cite{Maxwell1864}, in which one predicts the flexibility of ensembles formed by rigid bodies connected by flexible linkages \cite{Grimm1975, Hammonds1996, Broedersz2011, Sharma2016}. Frames -- consisting of rigid elements (\emph{sites}) and the connections between them (\emph{bonds}) \cite{lubensky2015phonons} -- are said to be rigid when the distance between two points cannot be altered without smoothly changing the length of one or more connections, and are said to be elastic otherwise. 

Notably, even in rigid frames, mechanical networks can undergo conformational changes that drastically alter their function, such as exotic shape transformations in metamaterials \cite{Bertoldi2017}, and allosteric regulation of enzymes where substrate binding in one region changes the structure and function of a distal active site \cite{Guo2016protein}. Characterizing and subsequently controlling such conformational changes is of critical import to a theoretical understanding of these materials, which in turn will support novel design and use of such materials. Yet, such characterization and control is challenged by the fact that perturbation to a few regions in the network can lead to wide-scale changes in the material's form in a complex manner that has to-date eluded formal treatment. Some have sought to address this challenge by designing networks through \emph{kinematic synthesis}, tracing arbitrary trajectories with a trace point using only a few actuators \cite{kempe1875trajectory, hartenberg1964linkage, kempe1877line}. Others have used computational heuristics such as tuning-by-pruning to predict mechanical responses in multiple nodes \cite{Rocks2017,goodrich2015pruning,Yan2017}. Given that such heuristics exist, it is now natural and timely to consider building a simple theory for how a mechanical network's topology constrains its control, and how novel topologies can be constructed to produce specified control functions.

Here we develop and exercise such a simple theory. We consider a rigid Maxwell Frame with nodes connected by edges. Given the connection topology, if we know all node positions, then the nullspace of the \emph{compatibility matrix} \cite{lubensky2015phonons} yields all allowed conformational changes. Alternatively, if we know all node displacements, then we can similarly find all node positions that yield that displacement. However, the successful characterization and control of designed networks must achieve desired positions and motions for a subset of the network, such as in auxetic materials that expand transversely in response to axial stretching.

We provide analytic and geometric principles for the construction, characterization, and control of these rigid frames with arbitrarily specified node positions and instantaneous displacements in $d$ dimensions. These principles are derived for bipartite frames, where edges are only allowed between (i) the subset of nodes whose positions and motions are specified and (ii) the subset of nodes whose positions and motions are unspecified. Building on prior work on the deformation of general and bipartite frames \cite{roth1981,whiteley1984,bolker1980}, we are able to characterize and control all networks that achieve arbitrarily and independently specified positions and motions in a subset of nodes. We are also able to characterize multi-purpose networks that achieve several distinct desired motions, and combine these networks as modules to construct large networks with desired conformational changes.

\section{Network Connectivity \& Mathematical Framework}

\begin{figure}[h!]
	\centering
	\includegraphics[width=0.4\columnwidth]{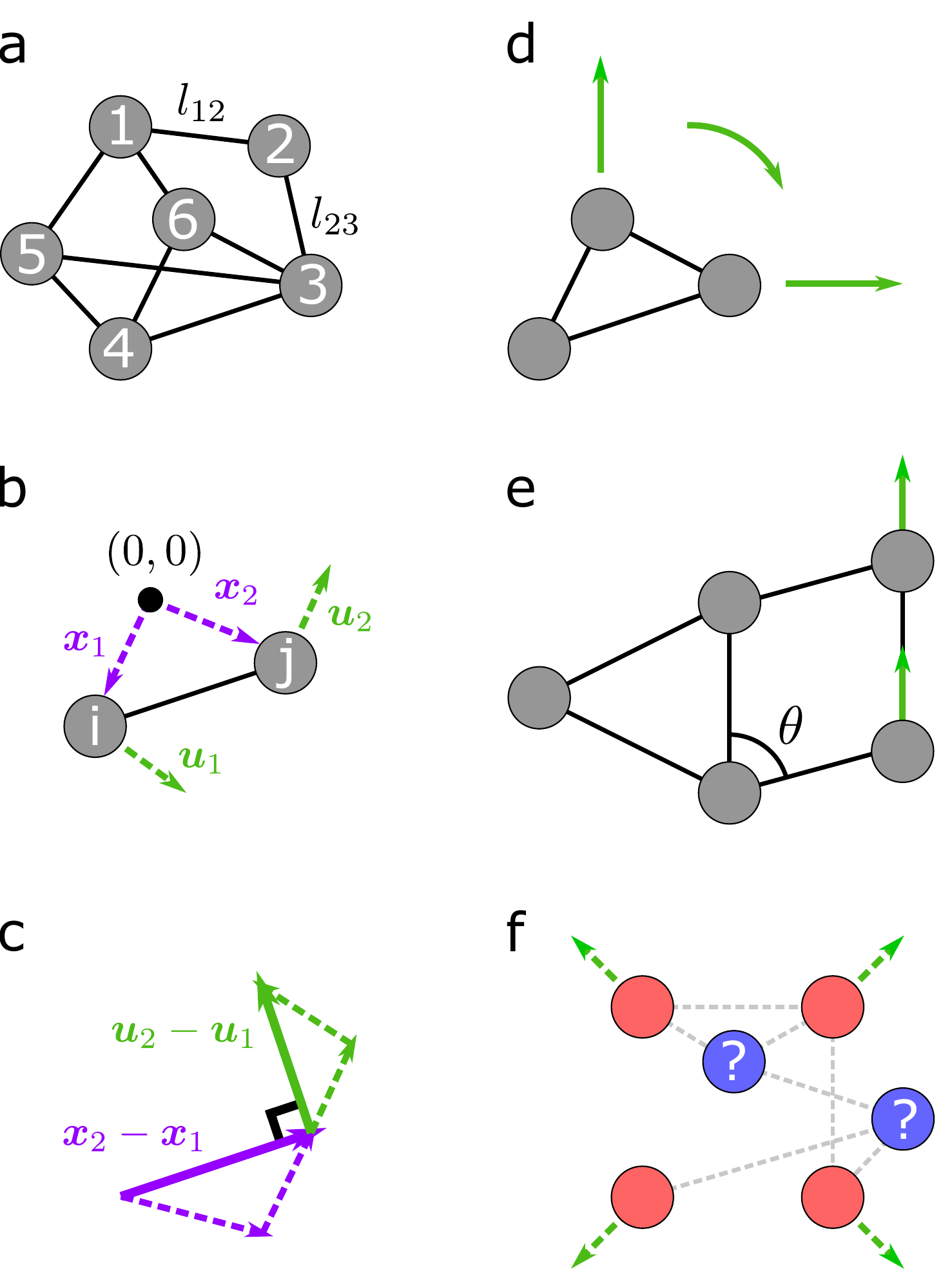}
	\caption{\textbf{Graphical Representations of Maxwell Frames.} (\textbf{a}) An example of a rigid frame in $d = 2$ dimensions with $N = 6$ nodes and $N_B = 9$ edges, marked with the length of the edge connecting node 1 to node 2, and the length of the edge connecting node 2 to node 3. (\textbf{b}) Two nodes connected by one edge, with the position vectors from an arbitrary origin specified in purple, and allowed displacement vectors in green, with (\textbf{c}) a graphical representation of the orthogonality of position and displacement vectors satisfying Eq.~\ref{eq:constraint}. (\textbf{d}) Graph of a rigid three-node system with the three rigid body motions in green. (\textbf{e}) Graph of a non-rigid five-node system with the fourth non-rigid body motion shown with green arrows, and parameterized by the continuous variable $\theta$. (\textbf{f}) Graph of four red specified nodes with desired displacements $\bm{u}_S$ shown with green arrows, potential edges in gray dashed lines, and unspecified nodes in blue.}
	\label{fig:mn_maxwell_frames}
\end{figure}

Consider the frame specified by the undirected graph $\mc G = (\mc V, \mc E)$ with $N$ nodes $\mc V = \{1,\dotsm,N\}$ embedded in $d$-dimensional space, and where the position of any node $i$ is specified by vector $\bm{x}_i \in \real^{d\times1}$. Further, consider a set of $N_B$ rigid edges $\mc E \subseteq \mc V \times \mc V$, where $l_{ij} = \|\bm{x}_i-\bm{x}_j\|_2$ is the length of the edge constraining the node pair $i,j$ (Fig.~\ref{fig:mn_maxwell_frames}a). For each node $i$, we specify an instantaneous motion with vector $\bm{u}_i \in \real^{d\times1}$. Then for any two nodes $i,j$ connected by an edge of length $l_{ij}$ (Fig.~\ref{fig:mn_maxwell_frames}b), the set of allowed motions $\bm{u}_i, \bm{u}_j$ must not change the length of $l_{ij}$ (Fig.~\ref{fig:mn_maxwell_frames}c). Hence, if an edge exists between node $i$ and node $j$, we require Eq.~\ref{eq:constraint} to linear order (see Supplemental Methods for derivation)
\begin{align}
\label{eq:constraint}
<\bm{x}_j - \bm{x}_i, \bm{u}_j - \bm{u}_i> = (\bm{x}_j - \bm{x_i})^T (\bm{u}_j - \bm{u}_i) = 0.
\end{align}

Further, our system of $N$ nodes has a corresponding set of $dN$ state variables, where each edge $l_{ij}$ provides a nonlinear distance constraint between two nodes. Provided that there are no states of self stress \cite{guest2006stiffness}, which we will ensure in the remainder of our results, the number of available finitely conformable degrees of freedom $N_D$ is given \cite{asimow1978rigidity} by Eq.~\ref{eq:dof}
\begin{align}
\label{eq:dof}
N_D = d\cdot N - N_B,
\end{align}
where $d(d+1)/2$ of these degrees of freedom are rigid body motions of translations and rotations that preserve distances between all pairs of nodes.

As a simple 2D example, we consider a triangle (Fig.~\ref{fig:mn_maxwell_frames}d) with $d = 2$, $N = 3$, $N_B = 3$, such that $N_D = 6-3 = 3$. We see that these three degrees of freedom correspond to the $x$-translation, $y$-translation, and rotation, and that the frame's configuration is fully determined by fixing 3 non-redundant $x$ or $y$ coordinates. Next we consider a more complex network of 5 nodes (Fig.~\ref{fig:mn_maxwell_frames}e), with $d=2, N=5, N_B = 6$, such that $N_D = 10-6=4$. Three of these degrees of freedom are rigid body motions, but the fourth manifests as a conformational deformation parameterized by $\theta$, which requires the setting of an additional fourth coordinate.

Given a frame $\mc G$ and given all node positions $\bm{x}_i$, these degrees of freedom are all instantaneous motions $\bm{u}_i$ that preserve all edge lengths according to Eq.~\ref{eq:constraint}. We can rewrite each constraint in terms of $\bm{u}_i$, and combine them to define the \emph{compatibility matrix} $C\in\real^{N_B\times dN}$ and displacement column vector $\bm{u} = [\bm{u}_1; \bm{u}_2; \dotsm; \bm{u}_N] \in \real^{dN \times 1}$ such that
\begin{align}
\label{eq:compatibility}
C\bm{u}=\bm{0},
\end{align}
where the $k$-th row of $C$, corresponding to the $k$-th edge connecting nodes $i$ and $j$, has all zero entries except $(\bm{x}_i-\bm{x}_j)^T$ multiplied by $\bm{u}_i$, and $(\bm{x}_j-\bm{x}_i)^T$ multiplied by $\bm{u}_j$. Then all node displacements $\bm{u}$ are given by the nullspace $\mc N(C)$.

In pursuing the understanding and control of mechanical materials, we are often interested in both the positions $\bm{x}_S \in \real^{dN_S \times 1}$ and displacements $\bm{u}_S \in \real^{dN_S \times 1}$ of a subset of network nodes (which we will call the \emph{specified nodes}) $\mc V_S \subset \mc V$ with $|\mc V_S| = N_S$ nodes, but not those $\bm{x}_U \in \real^{dN_U \times 1}$,  $\bm{u}_U \in \real^{dN_U \times 1}$ of the remaining nodes (which we will call the \emph{unspecified nodes}) $\mc V_U \subset \mc V$ with $|\mc V_U| = N_U$ nodes, such that $\mc V_S \cup \mc V_U = \mc V$, and $\mc V_S \cap \mc V_U = \emptyset$. Consideration of a subset of specified nodes is common in the study of several materials, such as those that have a negative Poisson ratio (Fig.~\ref{fig:mn_maxwell_frames}f). In what follows, we provide analytic and geometric principles for characterizing all bipartite networks that satisfy position and displacement conditions for specified nodes. We then demonstrate how these principles can be used to design networks that generate these desired motions by controlling only a few nodes.

\section{Results}

\subsection{Conic Sections and Overlaps of Bipartite Networks}

\begin{figure}[h!]
	\centering
	\includegraphics[width=.95\columnwidth]{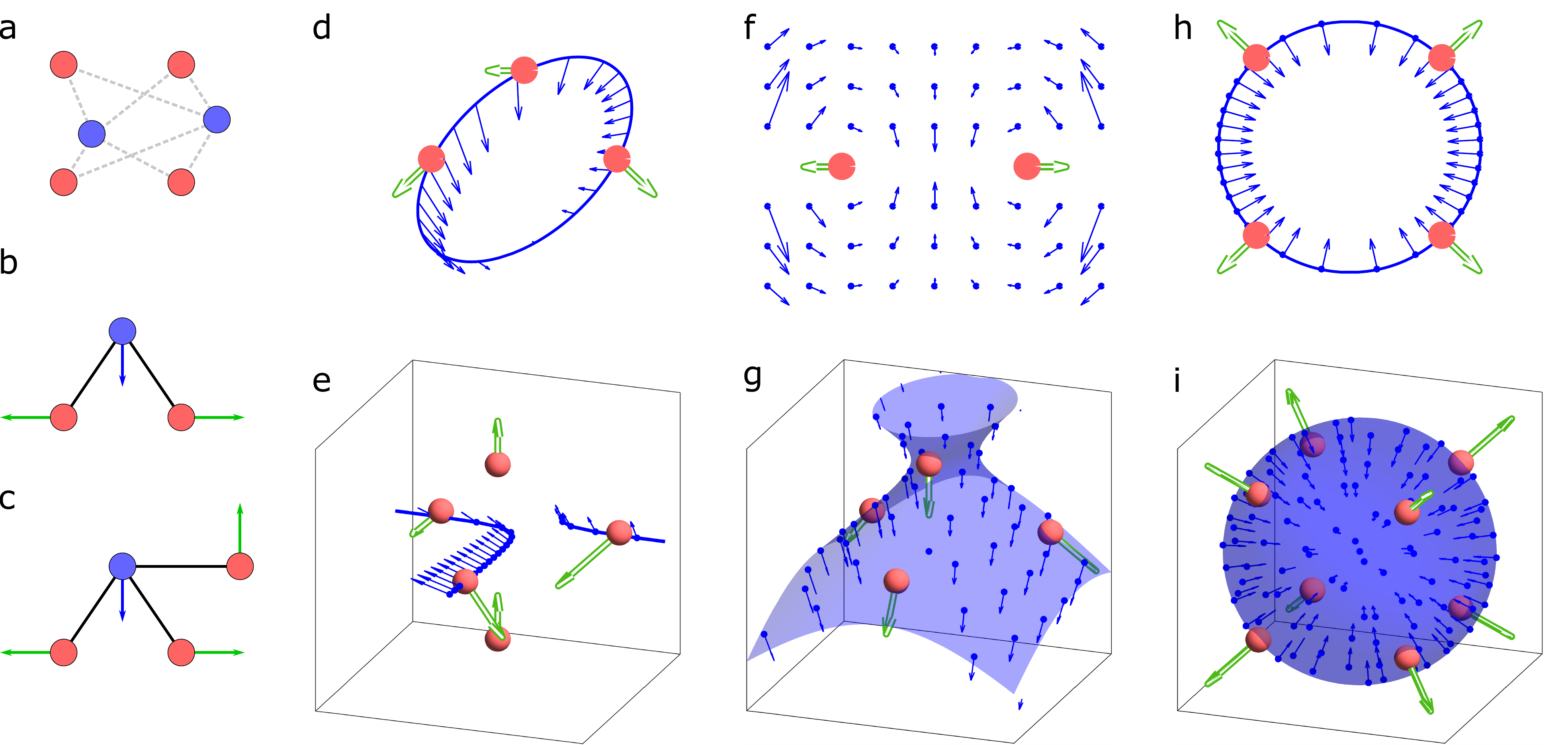}
	\caption{\textbf{Solution Space of Unspecified Nodes is Determined by the Position and Displacement of Specified Nodes.} (\textbf{a}) Example of a bipartite network with specified nodes in red, unspecified nodes in blue, and allowed edges in gray. (\textbf{b}) Position ($\bm{x}_{Uj}$, location of blue node) and displacement ($\bm{u}_{Uj}$, blue arrow) of an unspecified node $j$ connected to two specified nodes (red), and (\textbf{c}) three specified nodes (also red), with displacements $\bm{u}_S$ shown with green arrows. In both cases, the blue node and blue arrow represent one position $\bm{x}_{Uj}$ and displacement $\bm{u}_{Uj}$ satisfying Eq.~\ref{eq:constraint}. (\textbf{d}) One dimensional solution spaces of all possible positions (blue curve) and displacements (blue arrows) of an unspecified node (in blue) connected to the three specified nodes (in red) with specified motions (hollow green arrows) for $d = 2$, and (\textbf{e}) $d = 3$. (\textbf{f}) Two dimensional solution spaces of all possible unspecified node positions and displacements connected to the red nodes in $d = 2$, and (\textbf{g}) $d = 3$. (\textbf{h}) Solution space of unspecified node positions and motions, where the specified node positions and motions are redundant to create linear dependencies in the rows of $\tilde{A}$ to yield a larger than expected solution space for $d = 2$, and (\textbf{i}) $d = 3$.}
	\label{fig:mn_bipartite}
\end{figure}

Given a subset $\mc{V}_S$ of $k$ nodes with positions $\bm{x}_S$ and displacements $\bm{u}_S$ as specified constants, and a disjoint subset $\mc{V}_U$ of nodes with positions $\bm{x}_U$ and displacements $\bm{u}_U$ as unspecified variables, we consider bipartite frames where edges only exist between specified nodes and unspecified nodes such that $\mc E \subseteq \mc V_S \times \mc V_U$ (Fig.~\ref{fig:mn_bipartite}a). As examples in $d=2$, we show one position (blue node) and motion (blue arrow) of an unspecified node that satisfies edge constraints Eq.~\ref{eq:constraint} connected to two (Fig.~\ref{fig:mn_bipartite}b) and three (Fig.~\ref{fig:mn_bipartite}c) specified nodes. Here, for any unspecified node $j$ with positions $\bm{x}_{Uj}$ and motions $\bm{u}_{Uj}$ that is connected to all $k$ specified nodes, we find all $\bm{x}_{Uj}, \bm{u}_{Uj}$ that satisfy the edge constraints Eq.~\ref{eq:constraint} given fixed $\bm{x}_S, \bm{u}_S$. 

We begin by writing the $k$ edge constraints in terms of variables $\bm{x}_{Uj}, \bm{u}_{Uj}$
\begin{align}
\label{eq:constraint_nonlinear}
\underbrace{
\begin{bmatrix}
\bm{u}_{S1}^T & \bm{x}_{S1}^T\\
\vdots & \vdots\\
\bm{u}_{Sk}^T & \bm{x}_{Sk}^T
\end{bmatrix}
}_{A}
\underbrace{
\begin{bmatrix}
\bm{x}_{Uj} \\ 
\bm{u}_{Uj}
\end{bmatrix}
}_{\bm{v}}
=
\underbrace{
\begin{bmatrix} 
\bm{x}_{S1}^T\bm{u}_{S1}\\ 
\vdots\\
\bm{x}_{Sk}^T\bm{u}_{Sk}
\end{bmatrix}
}_{\bm{b}}
+
\underbrace{
\bm{x}_{Uj}^T\bm{u}_{Uj}
}_{c}
\underbrace{
\begin{bmatrix}
1\\
\vdots\\
1
\end{bmatrix}
}_{\bm{1}},
\end{align}
By temporarily omitting the constraint that $c = \bm{x}_{Uj}^T\bm{u}_{Uj}$, we can linearize Eq.~\ref{eq:constraint_nonlinear} to 
\begin{align}
\underbrace{
\begin{bmatrix}
A & -\bm{1}
\end{bmatrix}
}_{\tilde{A}}
\underbrace{
\begin{bmatrix}
\bm{v}\\
c
\end{bmatrix}
}_{\tilde{\bm{v}}}
=
\bm{b}.
\end{align}
Let the $m$-dimensional nullspace $\mc{N}(\tilde{A})$ be spanned by the basis set $W = [\bm{w}_1, \dotsm, \bm{w}_m]$. If $\bm{b} \in \mc R(\tilde{A})$, then a particular solution is given by $\tilde{\bm{v}}^* = \tilde{A}^+\bm{b}$, and the homogeneous solutions are given by linear combinations of the nullspace basis. Then $\tilde{\bm{v}} = \alpha_1\bm{w}_1 + \dotsm + \alpha_m\bm{w}_m + \tilde{\bm{v}}^* = \tilde{W}\bm{\tilde{\alpha}}$, where $\tilde{W} = \begin{bmatrix} \bm{w}_1, \dotsm, \bm{w}_m, \tilde{\bm{v}}^* \end{bmatrix}$ and $\bm{\tilde{\alpha}} = \begin{bmatrix}\alpha_1; & \dotsm; & \alpha_m; & 1\end{bmatrix}$. Finally, we apply the constraint $c = \bm{x}_{Uj}^T\bm{u}_{Uj}$ to yield all solutions $\bm{\tilde{\alpha}}$ to Eq.~\ref{eq:constraint_nonlinear} (see Supplementary Methods) by
\begin{align}
\label{eq:quadric}
\bm{\tilde{\alpha}}^T
Q
\bm{\tilde{\alpha}} = 0,
\end{align}
where $Q\in\real^{m+1 \times m+1}$. The solution space of $\bm{\tilde{\alpha}}$ to Eq.~\ref{eq:quadric} has dimension $m-1$, which are \emph{points} for $m=1$, \emph{conic sections} for $m=2$, and \emph{quadric surfaces} for $m=3$. 

For the general case where $\tilde{A}$ has full row rank, then $m = 2d+1-k$. In $d=2$, given $k=3$ independent specified node positions $\bm{x}_{S1}, \bm{x}_{S2}, \bm{x}_{S3}$ and motions $\bm{u}_{S1}, \bm{u}_{S2}, \bm{u}_{S3}$ such that $\dim(\mc N(\tilde{A})) = m=2$, the solution space for $\bm{x}_{Uj},\bm{u}_{Uj}$ is one dimensional (Fig.~\ref{fig:mn_bipartite}d). Similarly, in $d=3$ with $k=5$ specified nodes, we again have $\dim(\mc N(\tilde{A})) = 2$ for another one dimensional solution space (Fig.~\ref{fig:mn_bipartite}e). By reducing the number of nodes by one, we can increase $\dim(\mc N(\tilde{A})) = 3$ for a two dimensional solution space in $d=2$ (Fig.~\ref{fig:mn_bipartite}f) and $d=3$ (Fig.~\ref{fig:mn_bipartite}g). We can also achieve a higher dimensional solution space by creating redundancies in the specified node positions and motions such that the rows of $\tilde{A}$ are linearly dependent. For example, the $i$-th rows of $\tilde{A}$ given by $[\bm{u}_{Si}^T, \bm{x}_{Si}^T, -1]$ can be written as linear combinations of three vectors $\bm{v}_1 = [1, 1, 1, 1, -1]$, $\bm{v}_2 = [-1, 1, -1, 1, -1]$, $\bm{v}_3 = [1, -1, 1, -1, -1]$, such that $\bm{v}_4 = \bm{v}_2 + \bm{v}_3 - \bm{v}_1$ in $d=2$ (Fig.~\ref{fig:mn_bipartite}h) and four vectors in $d=3$ (Fig.~\ref{fig:mn_bipartite}i). Hence, these curves and surfaces characterize the only $\bm{x}_{Uj},\bm{u}_{Uj}$ of a node connected to all specified nodes that do not preclude the desired specified node positions and motions by construction.

\subsection{Network Construction Through Judicious Constraint Placement}

The quadrics and conics previously discussed specify all positions $\bm{x}_{Uj}$ and displacements $\bm{u}_{Uj}$ of an unspecified node $j$ that satisfy all $k$ edge constraints given the positions $\bm{x}_S$ and displacements $\bm{u}_S$ of $k$ specified nodes. Hence, the addition of each unspecified node along these surfaces adds $d$ state variables and $k$ constraints while preserving the desired motions $\bm{u}_S$. Here we demonstrate that for $k > d$, we can judiciously constrain our system such that the only non-rigid body degrees of freedom allows desired $\bm{x}_S, \bm{u}_S$, thereby requiring in general that the solution space have dimension of at least 0, and at most $d-1$ for $k = d+1$.

\begin{figure}[h!]
	\centering
	\includegraphics[width=0.95\columnwidth]{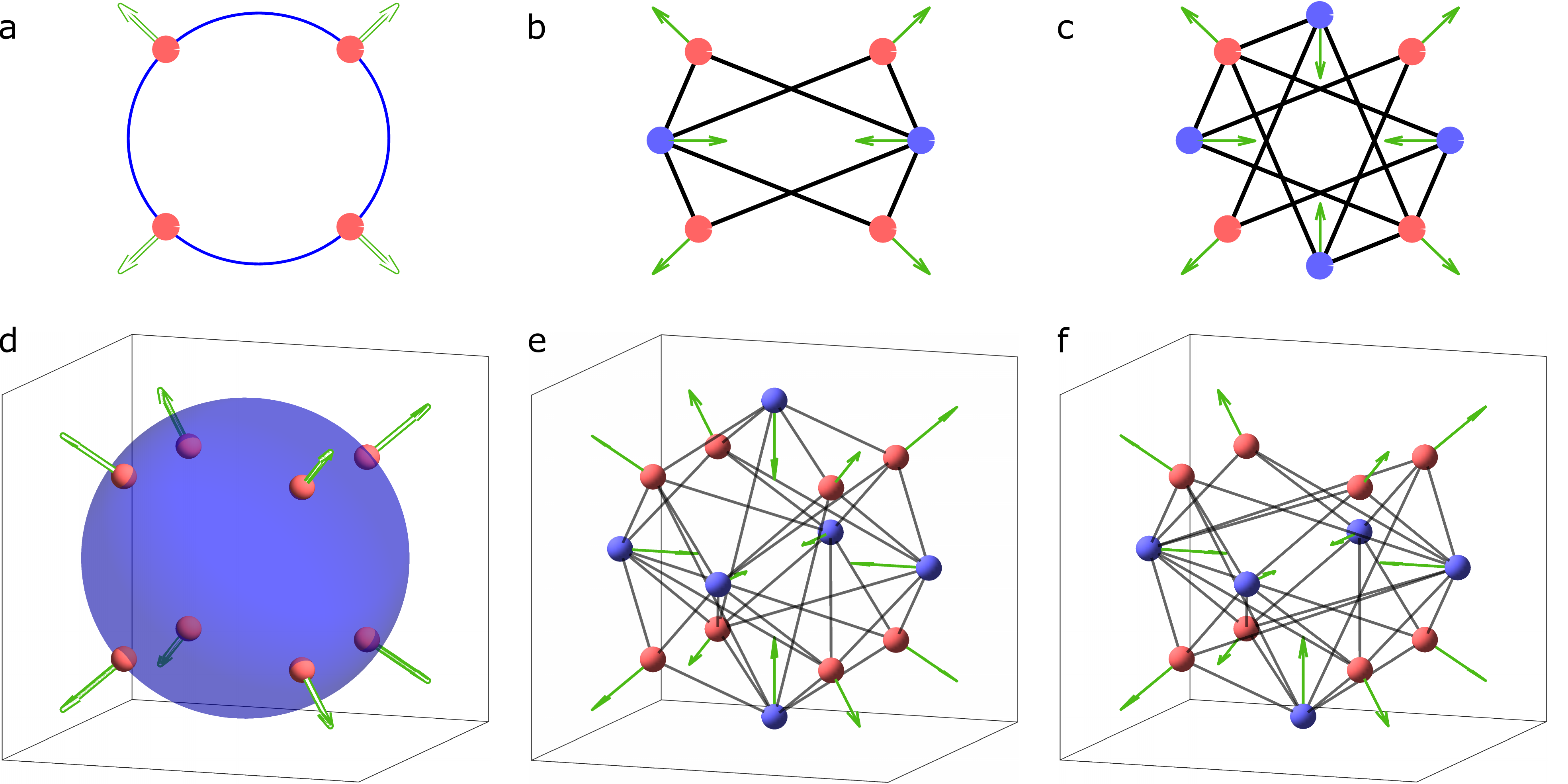}
	\caption{\textbf{Construction and Control of Frames with Specified Outward Motion.} (\textbf{a}) Schematic in $d = 2$ of four specified nodes with desired outward displacements (hollow green arrows) with the corresponding solution space of unspecified node positions (blue curve) satisfying Eq.~\ref{eq:constraint_nonlinear}. (\textbf{b}) Example bipartite frames with four degrees of freedom, constructed from placing 2 unspecified nodes, 8 edges, and (\textbf{c}) 4 unspecified nodes, 12 edges. The motion along the only non-rigid degree of freedom is given by the solid green arrows. (\textbf{d}) Schematic in $d=3$ of eight specified nodes with desired outward displacements (hollow green arrows), with a spherical unspecified node solution space (blue surface). (\textbf{e}) Bipartite frame with seven degrees of freedom, constructed by placing 5 unspecified nodes, 32 edges, and (\textbf{f}) 6 unspecified nodes, 35 edges, with the only non-rigid body motion shown with solid green arrows.}
	\label{fig:mn_construction}
\end{figure}

A system of $N_S$ disjoint specified nodes in $d$ dimensions has $N_D = dN_S$ degrees of freedom, of which $d(d+1)/2$ are rigid body. If we add $N_U$ unspecified nodes with $N_B$ edges following
\begin{align}
\label{eq:dof_bipartite}
N_D = d(N_S + N_U) - N_B = \frac{d(d+1)}{2}+1,
\end{align}
in such a way that 1) there are no states of self-stress, and 2) there are no rigid subgraphs, we ensure that our motion $\bm{u}_S$ is the only finitely deformable non-rigid degree of freedom. As an example in $d=2$, we provide an initial $N_S = 4$ node system with a desired outward motion (Fig.~\ref{fig:mn_construction}a) with a 1-dimensional solution space. Initially, the nodes have $N_D = dN_S = 8$, which we reduce to $4$ by adding $N_U = 2$ nodes with $N_B = 8$ edges (Fig.~\ref{fig:mn_construction}b), and $N_U = 4$ nodes with $N_B = 12$ edges (Fig.~\ref{fig:mn_construction}c) along the conic section to yield a single non-rigid motion. 

We can similarly achieve the same result for $N_S = 8$ specified nodes in $d = 3$ where $rank(\tilde{A}) = 4$ such that $\dim(\mc N(\tilde{A})) = 3$ for a 2-dimensional solution space (Fig.~\ref{fig:mn_construction}d) and desired $N_D = d(d+1)/2+1 = 7$. The desired motion exists as the single non-rigid motion after placing $N_U = 5$ nodes with $N_B = 32$ edges (Fig.~\ref{fig:mn_construction}e), or $N_U = 6$ nodes with $N_B = 35$ edges (Fig.~\ref{fig:mn_construction}f). Through the judicious addition of unspecified nodes with more edges than state variables ($k>d$) along our solution space, we reduce the network's degrees of freedom without precluding our specified node positions and motions.

\subsection{Multi-Mode Construction}

\begin{figure}[h!]
	\centering
	\includegraphics[width=0.95\columnwidth]{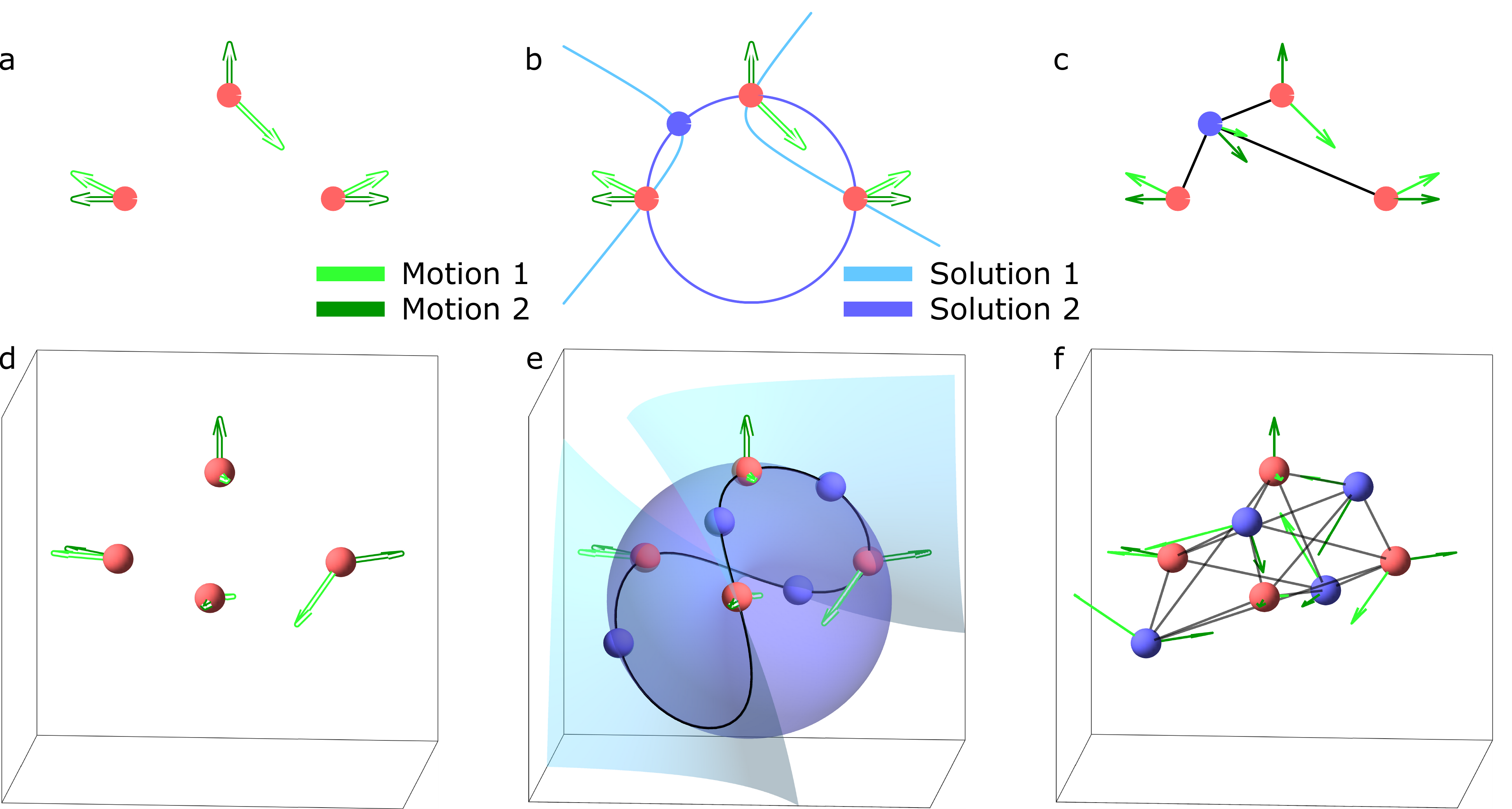}
	\caption{\textbf{Intersections of Unspecified Node Solution Spaces for Multiple Non-Rigid Motions.} (\textbf{a}) Schematic of three specified nodes with two sets of desired motions, $\bm{u}_{S1}$ (hollow light green arrows), and $\bm{u}_{S2}$ (hollow dark green arrows). (\textbf{b}) Solution spaces of unspecified node positions for desired motions $\bm{u}_{S1}$ (light blue curve) and for $\bm{u}_{S2}$ (dark blue curve), with one unspecified blue node at their intersection. (\textbf{c}) Constructed network with the two true non-rigid body degrees of freedom at all nodes shown with solid light and dark green arrows. (\textbf{d}) System of four specified nodes with two sets of desired motions in hollow light arrows and in dark green arrows. (\textbf{e}) Solution spaces of the unspecified node positions for motion 1 (light blue surface) and motion 2 (dark blue surface), with four unspecified blue nodes located at their intersection traced out by the black line. (\textbf{f}) Constructed network with the only two true non-rigid body degrees of freedom in all nodes shown in solid light arrows and in dark green arrows.}
	\label{fig:mn_multi}
\end{figure}

Using this judicious constraint principle, we characterize and create networks with specified positions $\bm{x}_S$ and two distinct displacements $\bm{u}_{S1}, \bm{u}_{S2}$. As an example in $d = 2$ for four nodes at $\bm{x}_S$ (Fig.~\ref{fig:mn_multi}a), we stipulate two motions $\bm{u}_{S1}$ and $\bm{u}_{S2}$, and we seek to characterize all bipartite networks that achieve these motions. 

We realize that for one set of specified positions $\bm{x}_{Si}$, each specified motion $\bm{u}_{Si}$ generates a solution space Eq.~\ref{eq:quadric} for the position and motion of an unspecified node $\bm{x}_{Uj}, \bm{u}_{Uj}$. For each quadratic $i$, this solution space has parameters $\bm{\tilde{\alpha}}_i$ that map to physical positions and motions based on $\tilde{W}$. We can define a projection from our solution coordinates $\tilde{\bm{\alpha}}$ to common spatial coordinates $\bm{\tilde{\beta}} = P_i \bm{\tilde{\alpha}}_i$. In this way, for $\tilde{Q}_i = P_i^{-T}Q_iP_i^{-1}$, solving for $\bm{\tilde{\beta}}$ in 
\begin{align}
\label{eq:basis}
\bm{\tilde{\beta}}^T \tilde{Q}_1 \bm{\tilde{\beta}} = 0,~~~~~\bm{\tilde{\beta}}^T \tilde{Q}_2 \bm{\tilde{\beta}} = 0,
\end{align}
finds the spatial coordinates where the solution spaces of both quadratics intersect. For $\dim(\mc N(\tilde{A})) = d$ with a solution dimension of $d-1$, the map $P_i$ is simply given by the first $d$ rows and columns of $\tilde{W}$. For $\dim(\mc{N}(\tilde{A})) = d-1$ with solution dimension $d-2$, the projection $\bm{\tilde{\beta}} = P_i \bm{\tilde{\alpha}}_i$ must also satisfy a linear constraint $\bm{p}^T P^{-1}\bm{\beta} = 1$ (see Supplementary Methods). While special cases in two equations and two variables permit an analytic expression for these intersections through the \emph{resultant} of the system \cite{macaulay1902elimination}, many numerical techniques also exist to find these solution space intersections in more than 2 variables. Hence, when designing 2 motions $\bm{u}_{S1}, \bm{u}_{S2}$, we can judiciously constrain our networks by placing our unspecified node at the intersection of the solution spaces while ensuring no states of self stress, and that no subgraph has fewer than 2 non-rigid body degrees of freedom to permit both motions.

To continue with our example in $d = 2$, we illustrate the solution spaces of both of these motions (Fig.~\ref{fig:mn_multi}b), and constrain our system to have $N_D = 5$, thereby allowing the two non-rigid body degrees of freedom (Fig.~\ref{fig:mn_multi}c). Similarly in $d=3$, we consider four specified nodes with two desired sets of motions (Fig.~\ref{fig:mn_multi}d), with solution spaces as surfaces (Fig.~\ref{fig:mn_multi}e). We observe that by placing our unspecified nodes along the line of the surface intersections, we can constrain our system to have $N_D = 8$ degrees of freedom (Fig.~\ref{fig:mn_multi}f), where the two true non-rigid body motions are identical to the desired motions. Through the placement of unspecified nodes along the intersection of the solution spaces for multiple desired motions, we judiciously constrain networks in $d=2$ and $d=3$ to preclude all but the two desired instantaneous motions as our two conformational degrees of freedom.

\subsection{Combining Network Modes for Potential Applications}

In the previous sections, we outlined how to judiciously constrain frames to allow for desired positions and motions in a subset of specified nodes. Here, we discuss a few basic approaches for combining these frames for a variety of potential applications. The core idea is to couple multiple bipartite modules each created with $d(d+1)/2 + 1$ degrees of freedom in a way that leaves the entire system with $N_D = d(d+1)/2 + 1$.

The natural way to couple these networks is to combine nodes. Consider the previously designed module in $d=2$ with $N_D = 4$ with outward motion (Fig.~\ref{fig:mn_construction}b). A system with two of these modules (Fig.~\ref{fig:mn_combine}a) has a total of $N_D = 8$. By merging two pairs of nodes between the modules, we remove two nodes corresponding to four state variables, bringing our system back to $N_d = 4$, compounding our motions (Fig.~\ref{fig:mn_combine}b). Due to this coupling, a conformational change in the top module requires a non-rigid body deformation in the bottom module. Hence, we require that for each module, the coupled nodes do not move as a rigid unit. For modules as these that preserve specific symmetries (see Supplementary Methods), we can create network lattices with properties such as negative Poisson ratio (Fig.~\ref{fig:mn_combine}c,d).

\begin{figure}[h!]
	\centering
	\includegraphics[width=1.0\columnwidth]{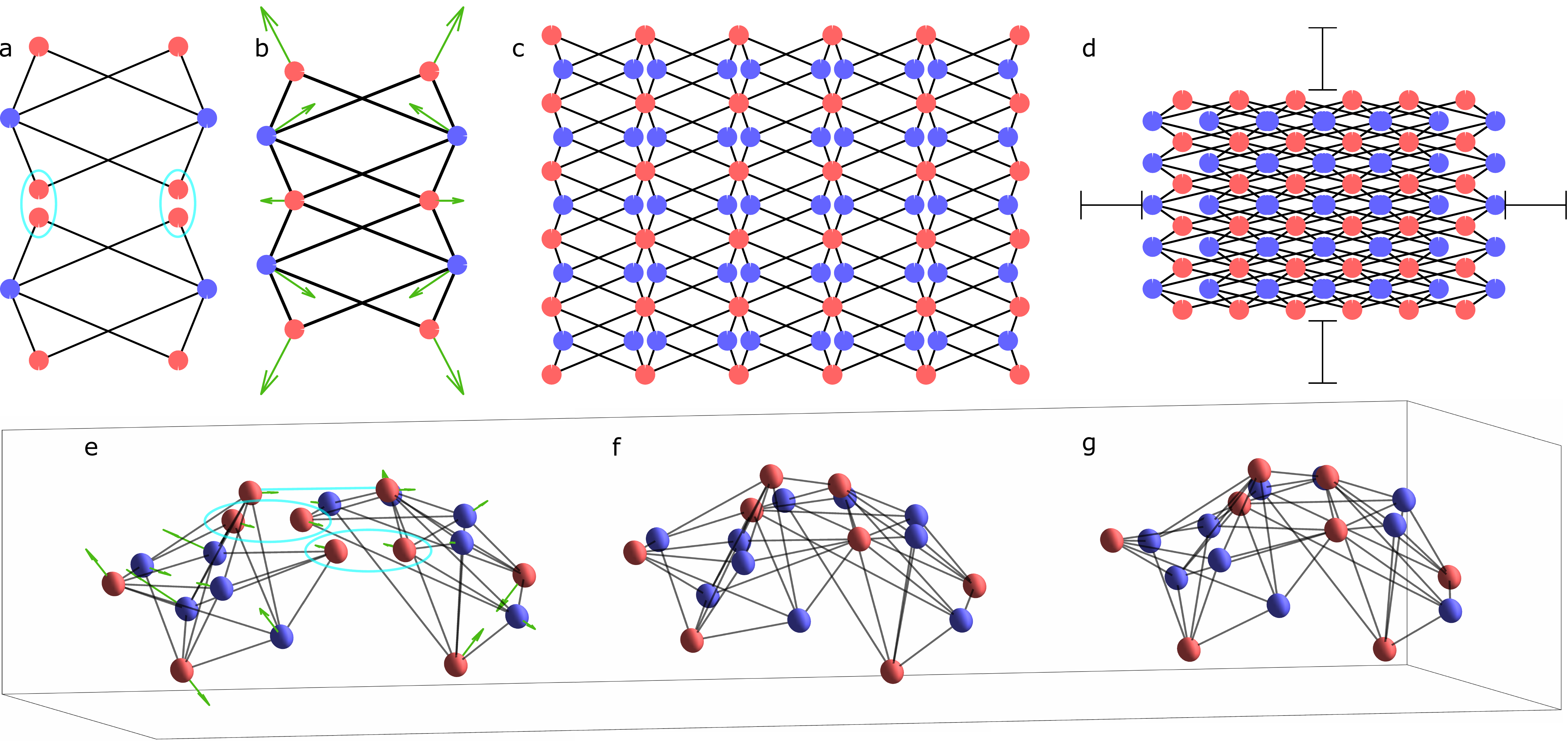}
	\caption{\textbf{Combining Network Motions by Merging Nodes and Adding Edges.} (\textbf{a}) Two independent outward moving modules in $d=2$ from Fig.~\ref{fig:mn_construction}b each with 4 degrees of freedom, where the pairs of nodes circled in cyan are to be merged. (\textbf{b}) Merged network with $N_D = 4$, with the one conformational degree of freedom in green arrows. (\textbf{c}) Large network of many coupled modules in the expanded and (\textbf{d}) contracted state, with bars to show the contraction distance from the expanded state. (\textbf{e}) Two independent modules in $d=3$, each with 7 degrees of freedom where the one conformational motion is shown in green arrows, with the nodes to be merged circled in cyan, and the edge to be added as a cyan line. (\textbf{f}) The combined network with one conformational motion in the contracted and (\textbf{g}) expanded state.}
	\label{fig:mn_combine}
\end{figure}

We can further remove degrees of freedom by adding extra bonds between module nodes. This bond addition becomes necessary in $d = 3$, as modules with one conformational motion have $N_D = d(d+1)/2+1 = 7$. Hence, two modules have $N_D = 14$, but coupling two pairs of nodes only removes 6 state variables instead of 7. For example, consider two modules (Fig.~\ref{fig:mn_combine}e) in $d=3$, each with one conformational degree for a total system $N_D = 14$. We can remove 6 degrees by coupling the two overlapping nodes (that also must not move as a rigid unit), and remove the last degree by adding an extra bond between the modules (Fig.~\ref{fig:mn_combine}f), to yield a coupled network with $N_D = 7$ that compounds our module motions (Fig.~\ref{fig:mn_combine}g). This long-range coupled conformational regulation of separately synthesized subunits is a hallmark of allostery in enzymes \cite{changeux1998} such as ATCase \cite{allewell1989,macol2001,cockrell2013}. Hence, by coupling modules through a combination of merging nodes (removes $d$ degrees) and adding extra bonds between nodes (removes 1 degree) that do not move as a rigid body, we can compound module motions. We address the construction of these modules and one more method of coupling using judicious constraint addition between modules in the Supplementary Methods.

\section{Discussion}

Deciphering principles of control in mechanical systems is of fundamental importance to understanding and optimizing the function of allosteric enzymes, auxetics, packings, and tunable metamaterials \cite{Bertoldi2017,Miskin2013adapting,jacobs2001protein,paulose2015topological,jacobs2012allostery}. Such principles could provide a link between network topology or geometry and the specific motions observed under controlling perturbations. In this work, we obtain an intuitive closed-form analytic solution space of the positions $\bm{x}_U$ and motions $\bm{u}_U$ of an unspecified node set $\mc{V}_U$ connected to a specified node set $\mc{V}_S$ to allow for the desired motions while reducing the total number of degrees of freedom. We further generalize this judicious constraint process to the design of two independent motions, and demonstrate how to design large network motions by coupling smaller modules. Taken together, our work provides fundamental analytic and geometric principles for the construction, characterization, and control of 2-D and 3-D mechanical Maxwell frames.

\noindent \textbf{Approaches to Network Design.} Important prior work in material design has focused on the use of algorithms to tune the responses of mechanical networks and packings \cite{Rocks2017,Miskin2013adapting,Yan2017,goodrich2015pruning}. For example, desired motions at multiple sites of an existing spring network can be tuned by the greedy iterative removal of bonds. In contrast to these prior studies, our approach can be used to provide a fully analytic characterization of all bipartite networks that achieve desired responses in an arbitrary number of nodes. In this sense, our characterization is complete, intuitive, and invariant to any algorithm, cost function, or initial condition of network topology and geometry. We also address the problem of characterizing the solution space of networks that simultaneously achieve multiple specified motions in response to different perturbations, which can be used for the design of networks with multiple functions. 

Important prior theoretical contributions provide valuable conditions for motions of bipartite frames \cite{whiteley1984}, or consider the properties and modification of predetermined structures, such as bistabilities of the Miura-ori tessellation \cite{silverberg2014origami}, and topological soft modes at dislocations in Kagome and square lattices \cite{paulose2015topological}. Prior work has also sought to systematically enumerate lattices that yield auxetic behavior \cite{korner2015auxetic}. Importantly, these works require a pre-existing structure, and do not address arbitrary and heterogeneous desired motions in the network. Our approach characterizes the full space of bipartite network solutions to achieve these arbitrary motions, which completely avoids complications associated with local minima and initial network configurations.

\noindent \textbf{Implications for Materials Physics and Mechanobiology.} Armed with the tools to generate desired motions in complex networks, we can begin thinking about applications in physics, biology, medicine, and engineering. One such application may be a bottom-up approach to designing cooperativity and allostery in proteins \cite{lukin2004heme,kamata2004allostery}, where we can use known protein structural motifs \cite{englander2014,papaleo2016} and simulations of protein tertiary structure to design macroscopic conformational changes. Another application is the design of materials with two independently controllable modes of deformation that behave auxetically or non-auxetically under different perturbations \cite{lee2012auxetic}. These tools could also be used to characterize solution spaces of networks with a specified motion, and to use these spaces as principled priors for the efficient search of materials with these desired properties \cite{dagdelen2017search}. Finally, we can design simple networks to generate precise and complex distributions of spatial forces using few actuators for complex biological tasks such as grasping in 3 dimensions \cite{yuzheng2005grasp}. In any application, we can develop a battery of modules that can be coupled to yield even more complex responses, simplifying the network design process into a module-coupling problem.

\noindent \textbf{Methodological Considerations.} Importantly, while we outlined conditions for successful network construction, ensuring no self stress is not always guaranteed. There exist pathological unspecified node placements along the intersection of multi-mode solution spaces, and along a plane in 3 dimensions that yield self stresses, some of which are explored \cite{whiteley1984}. 

Given the simplification of bipartite networks, these results only address the construction of more complex non-bipartite networks as the combination of bipartite modules. As such, an interesting future extension of this research would be the hierarchical judicious constraining of networks along the solution spaces in bipartite networks of specified \emph{and} unspecified nodes. Similarly, the analysis of existing networks is currently restricted to those that are well-approximated by this bipartite simplification. An interesting future extension here would be the deconstruction of existing non-bipartite networks into coupled bipartite modules. These motions are also designed in the linear regime, and as such cannot speak to the extent to which the nonlinear motion follows this linear approximation. Useful future work could use analytical and computational approaches to consider geometries along these solution spaces that yield robust nonlinear responses that can be coupled to produce robust bulk responses. Another useful future direction is the study of judicious constraint addition on non-bipartite networks, or the deconstruction of networks into interconnected modules of specified and unspecified nodes.

\section{Conclusion}
The simple and intuitive relationship between desired network motions and the full bipartite solution space is a powerful tool for the understanding and design of mechanical motions. These results open the door to a wide range of useful theoretical, computational, and experimental applications and extensions, including the hierarchical judicious constraint addition of non-bipartite networks, algorithms to deconstruct empirical networks into coupled bipartite modules, and the design of modules with robust nonlinear responses. From novel bottom-up design principles of protein mechanics to modular construction of bulk material properties, these tools can be applied and advanced in a wide range of activities. 

\section{Acknowledgments}
\noindent We gratefully acknowledge useful conversations with Bryan Chen, Ann Sizemore, and Eli Cornblath. JZK acknowledges support from National Institutes of Health T32-EB020087, PD: Felix W. Wehrli, and the National Science Foundation Graduate Research Fellowship No. DGE-1321851. SHS acknowledges support from the United States NSF Grant Nos. DMS-1513179 and CCF-1522054. DSB acknowledges support from the John D. and Catherine T. MacArthur Foundation, the ISI Foundation, the Alfred P. Sloan Foundation, an NSF CAREER award PHY-1554488, and from the NSF through the University of Pennsylvania Materials Research Science and Engineering Center (MRSEC) DMR-1720530.

\newpage
\section{Supplementary Methods}
\subsection{Key Formulations}

\subsubsection{Single Edge Constraint Between Nodes $i$ and $j$:}
\begin{align}
\label{eq:constraint}	
<\bm{x}_j - \bm{x}_i, \bm{u}_j - \bm{u}_i> = (\bm{x}_j - \bm{x}_i)^T (\bm{u}_j - \bm{u}_i) = 0.
\end{align}

\subsubsection{Degrees of Freedom of System of $N$ Nodes and $N_B$ Edges in $d$ Dimensions:}
\begin{align}
\label{eq:dof}
N_D = d \cdot N - N_B.
\end{align}

\subsubsection{Zero Mode Motions From Complimentary Matrix:}
\begin{align}
\label{eq:compatibility}
C\bm{u} = \bm{0}.
\end{align}

\subsubsection{Rewriting $k$ Edge Constraints from Specified Nodes $i = 1,\dotsm,k$ to One Unspecified Node $j$:}
\begin{align}
\label{eq:constraint_nonlinear}
\underbrace{
	\begin{bmatrix}
	\bm{u}_{S1}^T & \bm{x}_{S1}^T\\
	\vdots & \vdots\\
	\bm{u}_{Sk}^T & \bm{x}_{Sk}^T
	\end{bmatrix}
}_{A}
\underbrace{
	\begin{bmatrix}
	\bm{x}_{Uj} \\ 
	\bm{u}_{Uj}
	\end{bmatrix}
}_{\bm{v}}
=
\underbrace{
	\begin{bmatrix} 
	\bm{x}_{S1}^T\bm{u}_{S1}\\ 
	\vdots\\
	\bm{x}_{Sk}^T\bm{u}_{Sk}
	\end{bmatrix}
}_{\bm{b}}
+
\underbrace{
	\bm{x}_{Uj}^T\bm{u}_{Uj}
}_{c}
\underbrace{
	\begin{bmatrix}
	1\\
	\vdots\\
	1
	\end{bmatrix}
}_{\bm{1}}.
\end{align}

\subsubsection{Linear Representation of Solution Space Omitting Quadratic Constraint:}
\begin{align}
\label{eq:linear}
\underbrace{
	\begin{bmatrix}
	A & -\bm{1}
	\end{bmatrix}
}_{\tilde{A}}
\underbrace{
	\begin{bmatrix}
	\bm{v}\\
	c
	\end{bmatrix}
}_{\tilde{\bm{v}}}
=
\bm{b}.
\end{align}

\subsubsection{Satisfaction of Quadratic Constraint:}
\begin{align}
\label{eq:quadric}
\bm{\tilde{\alpha}}^T
Q
\bm{\tilde{\alpha}} = 0.
\end{align}

\subsubsection{Degrees of Freedom Given $N_U$ Unspecified Node Additions:}
\begin{align}
\label{eq:dof_bipartite}
N_D = d(N_S + N_U) - N_B = \frac{d(d+1)}{2}+1.
\end{align}

\subsubsection{Intersection of Two Solution Spaces for Multi-Mode Construction:}
\begin{align}
\label{eq:basis}
\bm{\beta}^T \tilde{Q}_1 \bm{\beta} = 0,~~~~~\bm{\beta}^T \tilde{Q}_2 \bm{\beta} = 0.
\end{align}

\subsection{Edge Constraint Satisfaction Through Perpendicular Relationship Between Node Positions and Motions}
Consider a network of $N$ nodes and $| \mc E | = N_B$ edges in $d$ dimensions. For any edge $e_k \in \mc E$ of length $l_k$ connecting node $i$ and node $j$, the node positions $\bm{x}_i = [x_{i1}; \dotsm; x_{id}], \bm{x}_j = [x_{j1}; \dotsm; x_{jd}]$ must satisfy the constraint 
\begin{align*}
g_k(\bm{x}) = (\bm{x}_i - \bm{x}_j)^T (\bm{x}_i - \bm{x}_j) = (x_{i1}-x_{j1})^2 + \dotsm + (x_{id}-x_{jd})^2 = l_k^2.
\end{align*}
We can gather all constraints $k = 1, \dotsm, N_B$ into an $N_B$ dimensional vector $\bm{g}(\bm{x})$
\begin{align*}
\bm{g}(\bm{x}) = 
\begin{bmatrix}
g_1(\bm{x})\\
g_2(\bm{x})\\
\vdots\\
g_{N_B}(\bm{x})\\
\end{bmatrix}
=
\begin{bmatrix}
l_1^2\\
l_2^2\\
\vdots\\
l_{N_B}^2
\end{bmatrix},
\end{align*}
and take the gradient with respect to instantaneous changes in node positions to get the \emph{Complementary Matrix} $C \in \real^{N_B \times dN}$
\begin{align*}
C = \nabla_{\bm{x}} \bm{g}(\bm{x}) =
\begin{bmatrix}
\frac{\partial g_1(\bm{x})}{\partial x_{11}}, & \dotsm, & \frac{\partial g_1(\bm{x})}{\partial x_{1d}}, & \frac{\partial g_1(\bm{x})}{\partial x_{21}}, & \dotsm, & \frac{\partial g_1(\bm{x})}{\partial x_{2d}}, & \dotsm, & \frac{\partial g_1(\bm{x})}{\partial x_{Nd}}\\
\frac{\partial g_2(\bm{x})}{\partial x_{11}}, & \dotsm, & \frac{\partial g_2(\bm{x})}{\partial x_{1d}}, & \frac{\partial g_2(\bm{x})}{\partial x_{21}}, & \dotsm, & \frac{\partial g_2(\bm{x})}{\partial x_{2d}}, & \dotsm, & \frac{\partial g_2(\bm{x})}{\partial x_{Nd}}\\
\vdots, & \ddots, & \vdots, & \vdots, & \ddots, & \vdots, & \dots, & \vdots\\
\frac{\partial g_{N_B}(\bm{x})}{\partial x_{11}}, & \dotsm, & \frac{\partial g_{N_B}(\bm{x})}{\partial x_{1d}}, & \frac{\partial g_{N_B}(\bm{x})}{\partial x_{21}}, & \dotsm, & \frac{\partial g_{N_B}(\bm{x})}{\partial x_{2d}}, & \dotsm, & \frac{\partial g_{N_B}(\bm{x})}{\partial x_{Nd}}
\end{bmatrix},
\end{align*}
which for any instantaneous motion $\bm{u} = [\bm{u}_1; \dotsm; \bm{u}_N] = [dx_{11}; \dotsm,; dx_{1d}; dx_{21}; \dotsm; dx_{Nd}]$, must not change the value of any constraint
\begin{align*}
C\bm{u} = \bm{0}.
\end{align*}
Hence, the requirement Eq.~\ref{eq:constraint} for edge $k$ comes from the fact that 
\begin{align*}
\nabla_{\bm{x}} g_k(\bm{x}) \bm{u}
&= 2(\bm{x}_i - \bm{x}_j)^T d\bm{x}_i - 2(\bm{x}_i - \bm{x}_j)^T d\bm{x}_j\\
&= 2(\bm{x}_i - \bm{x}_j)^T \bm{u}_i - 2(\bm{x}_i - \bm{x}_j)^T \bm{u}_j\\
&= 2(\bm{x}_i - \bm{x}_j)^T(\bm{u}_i - \bm{u}_j)\\
&= 0.
\end{align*}
We note that the complementary matrix only tests zero modes to linear order, and that the condition $\bm{u} \in \mc N(C)$ is necessary but not sufficient for the actual motion $\bm{u}$ to be a finite deformation. Given a system has no states of self stress, the motions $\bm{u} \in \mc N(C)$ are finitely deformable.

\subsection{Rewriting the Linearized Edge Constraints into Vector Form}
Consider a system of $k$ specified nodes where the $i$-th node has some constant desired position $\bm{x}_{Si}$ and motion $\bm{u}_{Si}$, and all $k$ nodes are connected to an unspecified node $j$ with variable position $\bm{x}_{Uj}$ and motion $\bm{u}_{Uj}$ for a total of $k$ edge constraints. Each linearized edge constraint can be written as
\begin{align*}
(\bm{x}_{Si} - \bm{x}_{Uj})^T (\bm{u}_{Si} - \bm{u}_{Uj})
&= \bm{x}_{Si}^T \bm{u}_{Si} - \bm{x}_{Si}^T \bm{u}_{Uj} - \bm{x}_{Uj}^T \bm{u}_{Si} + \bm{x}_{Uj}^T \bm{u}_{Uj}\\
&= \bm{x}_{Si}^T \bm{u}_{Si} - \bm{x}_{Si}^T \bm{u}_{Uj} - \bm{u}_{Si}^T \bm{x}_{Uj}  + \bm{x}_{Uj}^T \bm{u}_{Uj}\\
&= \bm{x}_{Si}^T \bm{u}_{Si} - 
\begin{bmatrix}
\bm{u}_{Si}^T & \bm{x}_{Si}^T
\end{bmatrix}
\begin{bmatrix}
\bm{x}_{Uj}\\
\bm{u}_{Uj}
\end{bmatrix}
+ \bm{x}_{Uj}^T \bm{u}_{Uj}\\
&= 0. 
\end{align*}
We can then consider a vector of constraints $\bm{g}(\bm{x}) = [g_1(\bm{x}); \dotsm; g_k(\bm{x})]$ to get
\begin{align*}
\begin{bmatrix}
(\bm{x}_{S1} - \bm{x}_{Uj})^T (\bm{u}_{S1} - \bm{u}_{Uj})\\
\vdots\\
(\bm{x}_{Sk} - \bm{x}_{Uj})^T (\bm{u}_{Sk} - \bm{u}_{Uj})
\end{bmatrix}
=
\begin{bmatrix}
\bm{x}_{S1}^T \bm{u}_{S1}\\
\vdots\\
\bm{x}_{Sk}^T \bm{u}_{Sk}
\end{bmatrix}
-
\begin{bmatrix}
\bm{u}_{S1}^T & \bm{x}_{S1}^T\\
\vdots\\
\bm{u}_{Sk}^T & \bm{x}_{Sk}^T
\end{bmatrix}
\begin{bmatrix}
\bm{x}_{Uj}\\
\bm{u}_{Uj}
\end{bmatrix}
+
\bm{x}_{Uj}^T \bm{u}_{Uj}
\begin{bmatrix}
1\\
\vdots\\
1
\end{bmatrix}
=
\begin{bmatrix}
0\\
\vdots\\
0
\end{bmatrix},
\end{align*}
from which we get Eq.~\ref{eq:constraint_nonlinear}
\begin{align*}
\underbrace{
	\begin{bmatrix}
	\bm{u}_{S1}^T & \bm{x}_{S1}^T\\
	\vdots & \vdots\\
	\bm{u}_{Sk}^T & \bm{x}_{Sk}^T
	\end{bmatrix}
}_{A}
\underbrace{
	\begin{bmatrix}
	\bm{x}_{Uj} \\ 
	\bm{u}_{Uj}
	\end{bmatrix}
}_{\bm{v}}
=
\underbrace{
	\begin{bmatrix} 
	\bm{x}_{S1}^T\bm{u}_{S1}\\ 
	\vdots\\
	\bm{x}_{Sk}^T\bm{u}_{Sk}
	\end{bmatrix}
}_{\bm{b}}
+
\underbrace{
	\bm{x}_{Uj}^T\bm{u}_{Uj}
}_{c}
\underbrace{
	\begin{bmatrix}
	1\\
	\vdots\\
	1
	\end{bmatrix}
}_{\bm{1}}.
\end{align*}

\subsection{Rewriting the Vector Form into a Linear Solution with Quadratic Constraint}
From Eq.~\ref{eq:linear}
\begin{align*}
\underbrace{
	\begin{bmatrix}
	A & -\bm{1}
	\end{bmatrix}
}_{\tilde{A}}
\underbrace{
	\begin{bmatrix}
	\bm{v}\\
	c
	\end{bmatrix}
}_{\tilde{\bm{v}}}
=
\bm{b},
\end{align*}
we know that if $\bm{b} \in \mc R(\tilde{A})$, one solution for the variables $\bm{\tilde{v}}^* = \tilde{A}^+ \bm{b}$, where $\tilde{A}^+$ is the Moore-Penrose pseudo-inverse. The full space of solutions $\tilde{\bm{v}}$ is given by the addition of $\bm{\tilde{v}}^*$ and any linear combination of vectors in the nullspace of $\tilde{A}$. Let ${\bm{w}_1, \dotsm, \bm{w}_m}$ be linearly independent vectors that span $\mc N(\tilde{A})$. Then we can construct $W = [\bm{w}_1, \dotsm, \bm{w}_m]$ with coordinates $\bm{\alpha} = [\alpha_1; \dotsm; \alpha_m]$ such that the solutions are given by 
\begin{align*}
\bm{\tilde{v}} = \alpha_1 \bm{w}_1 + \dotsm + \alpha_m \bm{w}_m + \bm{\tilde{v}}^* = 
\begin{bmatrix}
W & & \bm{\tilde{v}}^*
\end{bmatrix} 
\begin{bmatrix}
\bm{\alpha}\\
1
\end{bmatrix}
= \tilde{W} \tilde{\bm{\alpha}}.
\end{align*}
The application of the quadratic constraint $c = \bm{x}_{Uj}^T \bm{u}_{Uj}$ is achieved through some basic algebraic manipulation. Recall that $c$ is the first entry of $\bm{\tilde{v}}$. Hence, for vector $\bm{p} = [0; \dotsm; 0; 1]$, we see that
\begin{align*}
c = \bm{p}^T \bm{\tilde{v}}. 
\end{align*}
Next, we can define matrix $O$ 
\begin{align*}
O = \frac{1}{2}
\begin{bmatrix}
0_{d \times d} & I_{d \times d} & 0_{d \times 1}\\
I_{d \times d} & 0_{d \times d} & 0_{d \times 1} \\
0_{1 \times d} & 0_{1 \times d} & 0
\end{bmatrix},
\end{align*}
to extract the values of $\bm{x}_{Uj}$ and $\bm{u}_{Uj}$ from solution vector $\bm{\tilde{v}}$
\begin{align*}
\bm{x}_{Uj}^T \bm{u}_{Uj} = \bm{\tilde{v}}^T O \bm{\tilde{v}}.
\end{align*}
Hence, our quadratic constraint Eq.~\ref{eq:quadric} comes from satisfying
\begin{align*}
c &= \bm{x}^T_{Uj} \bm{u}_{Uj}\\
\bm{p}^T \bm{\tilde{v}} &= \bm{\tilde{v}}^T O \bm{\tilde{v}}\\
\bm{p}^T
\begin{bmatrix}
W & \bm{\tilde{v}}^*
\end{bmatrix} 
\begin{bmatrix}
\bm{\alpha}\\
1
\end{bmatrix}
&= 
\begin{bmatrix}
\bm{\alpha}^T & 1
\end{bmatrix} 
\begin{bmatrix}
W^T\\
\bm{\tilde{v}}^{*T}
\end{bmatrix}
O
\begin{bmatrix}
W & & \bm{\tilde{v}}^*
\end{bmatrix} 
\begin{bmatrix}
\bm{\alpha}\\
1
\end{bmatrix}\\
\bm{p}^T W \bm{\alpha} + \bm{p}^T \bm{\tilde{v}}^*
&=
\bm{\alpha}^T W^T O W \bm{\alpha} + 2\bm{\tilde{v}}^{*T}OW\bm{\alpha} + \bm{\tilde{v}}^{*T}O\bm{\tilde{v}}^*,
\end{align*}
and we can group the variables $\bm{\alpha}$ to get the form
\begin{align*}
\bm{\alpha}^T [W^TOW] \bm{\alpha} + [2\bm{\tilde{v}}^{*T}O - \bm{p}^T]W\bm{\alpha} + [\bm{\tilde{v}}^{*T}O - \bm{p}^T]\bm{\tilde{v}}^* = 0.
\end{align*}
To make the equation more presentable, we define $A = [W^TOW]$, $B = [2\bm{\tilde{v}}^{*T}O - \bm{p}^T]W/2$, and $C = [\bm{\tilde{v}}^{*T}O - \bm{p}^T]\bm{\tilde{v}}^*$ to write
\begin{align*}
Q = 
\begin{bmatrix}
A & B\\
B^T & C
\end{bmatrix},
\end{align*}
such that the quadratic constraint $c = \bm{x}_{Uj}^T\bm{u}_{Uj}$ is rewritten
\begin{align*}
\begin{bmatrix}
\bm{\alpha}^T & 1
\end{bmatrix}
Q
\begin{bmatrix}
\bm{\alpha}\\
1
\end{bmatrix}
=
0.
\end{align*}

\subsection{Dimensionality of the Unspecified Solution Space}
The dimensionality of the positions $\bm{x}_{Uj}$ and motions $\bm{u}_{Uj}$ of an unspecified node is completely determined by the dimension of $\mc N(\tilde{A})$. Because the linear solutions of $\bm{\tilde{v}}$ have dimension $\dim(\mc N(\tilde{A}))$, and we have only one more constraint Eq.~\ref{eq:quadric}, the dimensionality of the unspecified node solution space is simply $\dim(\mc N(\tilde{A})) - 1$.

\subsection{Symmetry Preservation in Determining Solution Space Dimension}
Given a system of $N_S$ specified nodes where all node positions $\bm{x}_{Si}$ and motions $\bm{u}_{S1}$ are linearly independent, the solution space dimension is generally given by $2d - N_S$. This result follows simply from the fact that $\tilde{A} \in \real^{N_S \times 2d+1}$, such that if all rows $i$ of $\tilde{A}$ given by $[\bm{x}_{Si}^T,~~ \bm{u}_{Si}^T,~~ -1]$ are linearly independent, then $\tilde{A}$ has full row rank, and $\dim(\mc N(\tilde{A})) = 2d+1 - N_S$. Finally, as a result of the quadratic constraint Eq.~\ref{eq:quadric}, the total solution dimension decreases by one to yield $2d - N_S$. If our specified node positions and motions are linearly dependent, then for each vector that is linearly dependent, the dimension of the solution space increases by one. Importantly, we note that these calculations are completely predicated on whether our specified node positions and motions are in the columnspace, such that $\bm{b} \in \mc R(\tilde{A})$. If $\bm{b} \notin \mc R(\tilde{A})$, then no solution exists.

\subsection{Defining a Projection from Solution Coordinates to Spatial Coordinates}
Our solution space for unspecified node positions and motions is given by linear combinations $\bm{\tilde{v}} = \tilde{V} \bm{\tilde{\alpha}}$ constrained by $\bm{\tilde{\alpha}}^T Q \bm{\tilde{\alpha}} = 0$. However, when solving for the intersection of solutions for the design of multiple motions $\bm{u}_{S1}, \bm{u}_{S2}$, we have multiple matrices $\tilde{A}_1, \tilde{A}_2$, where the variables $\bm{\tilde{\alpha}}_1, \bm{\tilde{\alpha}}_2$ are not necessarily represented in the same spatial coordinates. To meaningfully solve for these intersections, we must first transform our solution coordinates $\bm{\tilde{\alpha}}$ into spatially meaningful coordinates in $d$ dimensions (e.g., $x, y, z$). A crucial component of this transformation is the dimension of the coordinate space, given by $\dim(\mc N(\tilde{A})) = m$.

\subsubsection{Case 1: Number of Solution Coordinates: $m = d$} 
If $m = d$, then we have at least $d$ solution coordinates in $\bm{\tilde{\alpha}}$. To convert to spatially meaningful coordinates, we seek a transformation matrix $P$ such that $\tilde{\bm{\beta}} = P \tilde{\bm{\alpha}}$. We desire that the $d$ entries of $\bm{\tilde{\beta}}$ correspond to spatial coordinates (e.g., $x, y, z$ for $d = 3$). Recall that
\begin{align*}
\bm{\tilde{v}} 
=
\begin{bmatrix}
\bm{x}_{Uj}\\
\bm{u}_{Uj}\\
c
\end{bmatrix}
= 
\begin{bmatrix}
W & \bm{\tilde{v}}^*
\end{bmatrix}
\begin{bmatrix}
\bm{\alpha}\\
1
\end{bmatrix},
\end{align*}
such that linear combinations of the first to $d$ rows of $\tilde{W}$ correspond to the spatial coordinates we seek as the first $d$ entries of $\tilde{\beta}$. Specifically, we can write
\begin{align*}
B = 
\begin{bmatrix}
I_{m \times m} & 0_{m \times 1} & 0_{m \times 2d-m}
\end{bmatrix},
\end{align*}
as a matrix that isolates the first to $m=d$ rows of $\tilde{W}$ via multiplication $B\tilde{W}$, to yield
\begin{align*}
\bm{\tilde{\beta}} = 
\begin{bmatrix}
\bm{\beta}\\
1
\end{bmatrix}
= 
\begin{bmatrix}
BW & B\bm{\tilde{v}}^*\\
0_{1 \times m} & 1
\end{bmatrix}
\bm{\tilde{\alpha}}
=
P \bm{\tilde{\alpha}}.
\end{align*}
With this transformation, we can create a transformed quadratic form
\begin{align*}
\bm{\tilde{\alpha}}^T Q \bm{\tilde{\alpha}} = \bm{\tilde{\beta}}^T \tilde{P}^{-T} Q \tilde{P}^{-1} \bm{\tilde{\beta}} = \bm{\tilde{\beta}}^T \tilde{Q}\bm{\tilde{\beta}} = 0,
\end{align*}
where the first $d$ entries of the solution $\bm{\tilde{\beta}}$ will be in spatial coordinates. 

\subsubsection{Case 2: Number of Solution Coordinates: $m = d-1$}
If the number of original coordinates is $m = d-1$, then we have one fewer solution dimensions than spatial dimensions, and a direct linear transformation matrix $P$ is insufficient. We move forward by treating the particular solution $\bm{\tilde{v}}^*$ as a part of the homogeneous solution in $\mc N(\tilde{A})$ such that
\begin{align*}
\bm{\tilde{v}} 
=
\begin{bmatrix}
c\\
\bm{x}_{Uj}\\
\bm{u}_{Uj}
\end{bmatrix}
= 
\begin{bmatrix}
W & \bm{\tilde{v}}^*
\end{bmatrix}
\begin{bmatrix}
\bm{\alpha}\\
\hat{\alpha}
\end{bmatrix},
\end{align*}
where $\hat{\alpha}$ should equal 1. Then similar to before, we select the spatial coordinates of our solution using matrix
\begin{align*}
B = 
\begin{bmatrix}
I_{d \times d} & 0_{d \times 1} & 0_{d \times d}
\end{bmatrix},
\end{align*}
to get the transformation
\begin{align*}
\bm{\tilde{\beta}} = 
\begin{bmatrix}
\bm{\beta}\\
\hat{\beta}
\end{bmatrix}
= 
B\begin{bmatrix}
W & \bm{\tilde{v}}^*
\end{bmatrix}
\begin{bmatrix}
\bm{\alpha}\\
\hat{\alpha}
\end{bmatrix}
=
P \bm{\tilde{\alpha}}.
\end{align*}
However, because $\hat{\alpha}$ must be 1, we have the extra constraint that for $\bm{p} \in \real^{d \times 1}$ where $\bm{p} = [0; \dotsm; 0; 1]$,
\begin{align*}
\bm{p}^T \bm{\tilde{\alpha}} = \hat{\alpha} = \bm{p}^T P^{-1}\bm{\tilde{\beta}} = 1.
\end{align*}
Hence, our transformation leads to the same form as the previous case
\begin{align*}
\bm{\tilde{\alpha}}^T Q \bm{\tilde{\alpha}} = \bm{\tilde{\beta}}^T P^{-T} Q P^{-1} \bm{\tilde{\beta}} = \bm{\tilde{\beta}}^T \tilde{Q}\bm{\tilde{\beta}} = 0,
\end{align*}
with the added condition that
\begin{align*}
\bm{p}^T P^{-1}\bm{\tilde{\beta}} = 1.
\end{align*}
Intuitively, what we have done is artificially extend our solution space to $d$ coordinates such that our quadratic constraint $\bm{\tilde{\beta}}^T \tilde{Q} \bm{\tilde{\beta}} = 0$ defines a $d - 1$ dimensional manifold, and we realize that the true solution space lies at the intersection of this manifold and the $d-1$ dimensional hyperplane defined by $\bm{p}^T P^{-1} \bm{\tilde{\beta}} = 1$. This way, we can change the coordinates of our original quadratic forms to $d$ spatial coordinates, and find the intersection of these quadratics and mathematically well defined hyperplanes.

\section{Combining Networks with Repeating Modules}
In the main text, we discuss how to couple two modules through judicious constraint addition. This method is the most general case, when the full nonlinear finite motions of the two modules do not overlap. Here, we will outline a simpler method of combining identical modules that share overlaps in their full nonlinear finite motions.

\begin{figure}[h!]
	\centering
	\includegraphics[width=1.0\columnwidth]{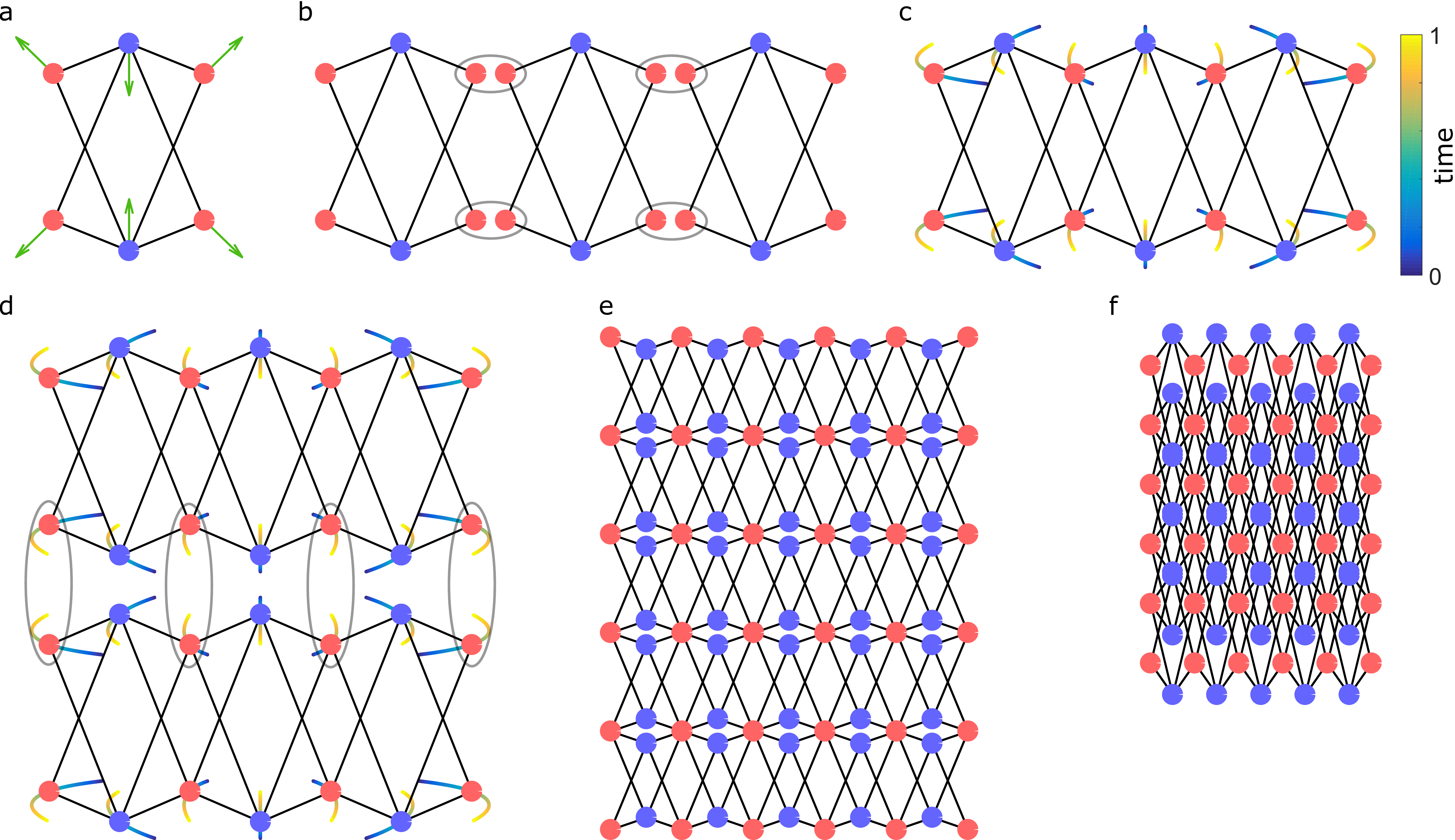}
	\caption{\textbf{Combination of Identical Modules with Nonlinear Symmetries Through Node Merging.} (\textbf{a}) Single module in 2 dimensions with 4 specified nodes (red), 2 unspecified nodes (blue), with one non-rigid body degree of freedom (green arrows). (\textbf{b}) Three replicates of the same module placed side-by-side, with the nodes to be merged grouped in the gray curves. (\textbf{c}) Nonlinear motion of combined network after merging grouped nodes, where the only non-rigid body motion is traced from blue to yellow for each node. (\textbf{d}) Two of these composite networks, aligned side-by-side with nodes to be merged grouped in the gray curves. (\textbf{e}) Full composite network with 4 horizontal and 4 vertical replicate modules with one non-rigid body degree of freedom in the expanded, and (\textbf{f}) contracted forms.}
	\label{fig:supp_combine_sym}
\end{figure}

To begin, we consider a simple module (Fig.~\ref{fig:supp_combine_sym}a) in 2 dimensions with 4 specified nodes, 2 unspecified nodes, and 1 non-rigid body degree of freedom shown in green arrows. Notice that this module has two symmetries: one along the horizontal axis, and one along the vertical axis. We can replicate this module along one of these directions (Fig.~\ref{fig:supp_combine_sym}b), and couple their specified nodes according to the gray curves. Note that this system of 3 modules has $4 \times 3 = 12$ degrees of freedom, and by grouping the two specified nodes into one node, we remove 4 nodes and $4 \times 2 = 8$ state variables to yield $12 - 8 = 4$ degrees of freedom. We show this composite network (Fig.~\ref{fig:supp_combine_sym}c), with the one non-rigid body degree of freedom shown in the full non-linear trajectory with curves for each node parameterized by a time variable from blue to yellow. 

We can then replicate this composite network (Fig.~\ref{fig:supp_combine_sym}d), and we notice that for the full non-linear conformational response, the grouped red specified node motions overlap. What we mean here is that as this time parameter is varied from 0 to 1, the x-coordinates of each grouped pair of nodes are equivalent, and the y-coordinates of each grouped pair of nodes is only offset by a single constant $c(t)$ across all groups, which is simply a rigid body translation. Alternatively, we can say that if we were to combine the node pairs in each group, we only require the addition of rigid body motions to one composite's nonlinear trajectory to exactly follow the other composite's trajectory in the grouped nodes. As an example, we replicate the single module in Fig.~\ref{fig:supp_combine_sym}a four times horizontally, and four times vertically, to create a networked sheet with one non-rigid body degree of freedom that we show in the expanded form (Fig.~\ref{fig:supp_combine_sym}e), and contracted form (Fig.~\ref{fig:supp_combine_sym}f). Hence, through the simple replicating and merging of simple modules that preserve certain symmetries, we can create materials that replicate the behavior of one module on a larger scale.

\section{Combining Networks Through Judicious Constraint Placement}
Consider a set of $|\mc V_S| = N_S$ specified nodes embedded in $d$ dimensions with coordinates $\bm{x}_S \in \real^{dN_S\times 1}$, and with desired displacements $\bm{u}_{S} \in \real^{dN_S \times 1}$. In general, the solution space of an unspecified node $j$ with $2d$ variables (position $\bm{x}_{Uj}$ and motion $\bm{u}_{Uj}$) constrained by connections to $k$ specified nodes has dimension $2d - k$. For $N_S > 2d$, we generally cannot place unspecified nodes connected to all $N_S$ specified nodes in a manner that preserves desired motions $\bm{u}_S$. 

\begin{figure}[h!]
	\centering
	\includegraphics[width=0.85\columnwidth]{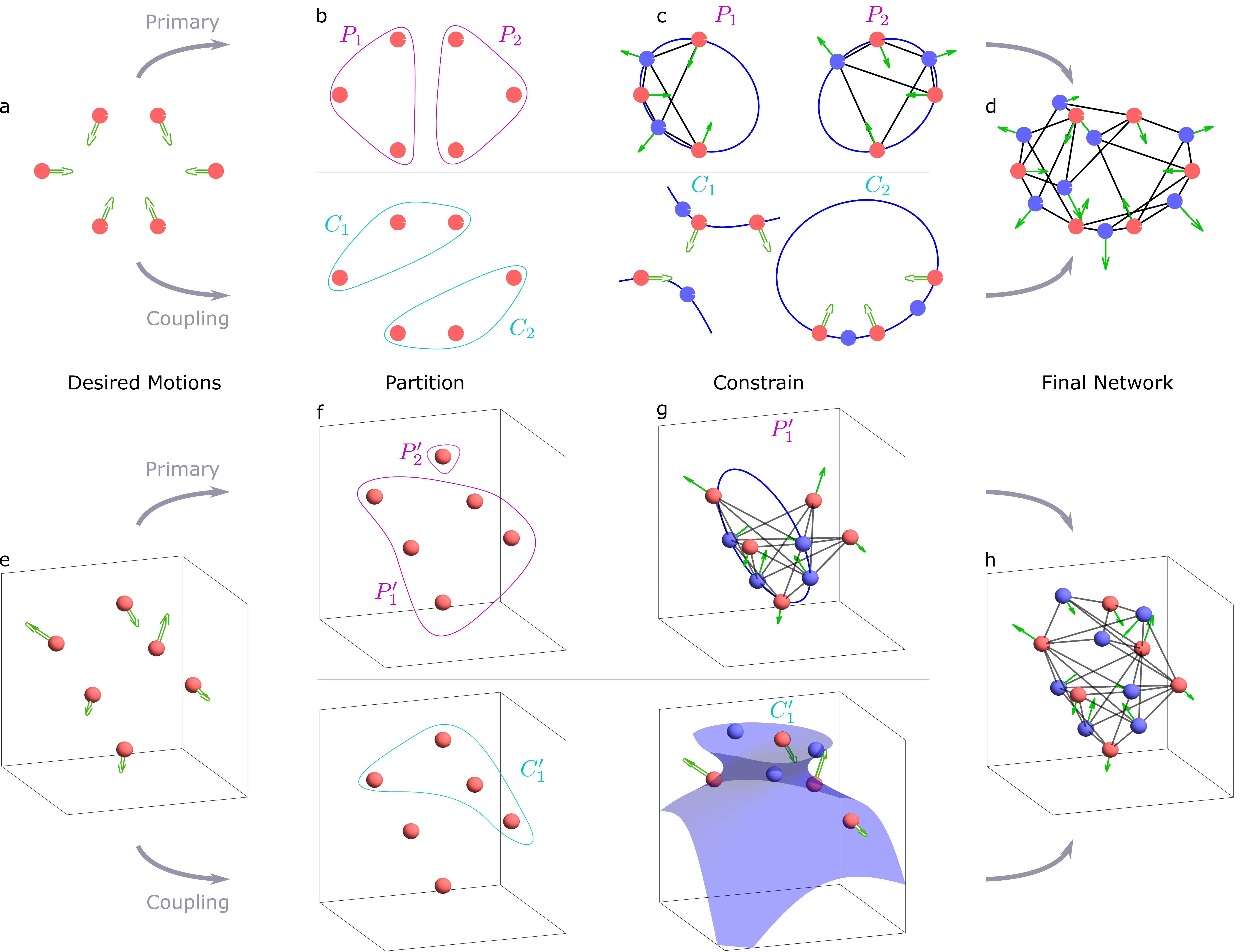}
	\caption{\textbf{Construction of Large Network Motions Through the Judicious Coupling Between Non-Intersecting Modules.} (\textbf{a}) Example in $d = 2$ of six specified node positions (red) and motions (hollow green arrows) with no solution for the placement of an unspecified node. (\textbf{b}) Partitioning of specified nodes into two primary modules ($P_1, P_2$, purple curve), coupled by two coupling modules ($C_1, C_2$, light blue curve). (\textbf{c}) Construction of primary modules $P_1, P_2$ through judicious constraint placement such that both primary modules have four degrees of freedom, with the non-rigid motion in solid green arrows, followed by the judicious constraining of the coupled modules $C_1, C_2$ by placing unspecified nodes (blue circles) on the solution space (blue curve) to bring $N_D$ of the total system (\textbf{d}) down to 4, with the only non-rigid body motion shown in solid green arrows. (\textbf{e}) Example in $d = 3$ of six specified nodes with no unspecified solution space, (\textbf{f}) partitioned into two primary modules ($P_1, P_2$, purple curve) and one coupling module ($C_1$, light blue curve). (\textbf{g}) Judicious constraint construction of primary module $P_1$ to seven degrees of freedom by placing four unspecified nodes along the unspecified solution space, and judicious constraint placement of the coupling module $C_1$ with a two dimensional solution space (blue surface) to yield (\textbf{h}) the constructed network with the true and only non-rigid body motion (solid green arrows).}
	\label{fig:mn_combine}
\end{figure}

Instead, we can partition the $N_S$ nodes into $p$ non-overlapping \emph{primary} modules $P_i \subseteq \mc V_S$ where $|P_i| \leq 2d$ nodes and $P_i \cap P_j = \emptyset$, and we judiciously constrain each module individually through the judicious placement of unspecified nodes to have $d(d+1)/2 + 1$ degrees of freedom. Then, we can couple these modules by constraining a second set of \emph{coupling} modules $C_k \subset \{P_i \cup P_j\}$ such that $|C_k| \leq 2d$, while ensuring the entire network has the necessary number of degrees of freedom to achieve the desired motion.

As an example in $d=2$, we partition a set of 6 specified nodes with desired motions (Fig.~\ref{fig:mn_combine}a) into two primary modules $P_1, P_2$ and two coupling modules $C_1, C_2$ (Fig.~\ref{fig:mn_combine}b) where $|P_1|=|P_2|=|C_1|=|C_2|=3$. We judiciously constrain the primary modules to have 4 degrees of freedom, and also judiciously constrain coupling modules $C_1, C_2$ (Fig.~\ref{fig:mn_combine}c) until the final network has 4 degrees of freedom, with our desired motion as the one non-rigid body motion (Fig.~\ref{fig:mn_combine}d). As another example in $d=3$, we partition a set of 6 specified nodes (Fig.~\ref{fig:mn_combine}e) into two primary modules $P_1', P_2'$ and one coupling module $C_1'$ (Fig.~\ref{fig:mn_combine}f), where $|P_1'| = 5$, $|P_2'| = 1$, and $|C_1| = 4$. We first judiciously constrain $P_1'$ along the unspecified solution space until it has $N_D = 7$; then we constrain the coupling module $C_1'$ (Fig.~\ref{fig:mn_combine}g) until the final network (Fig.~\ref{fig:mn_combine}h) has $N_D = 7$, with the desired motion as the only non-rigid body degree of freedom. We see that by judiciously constraining these primary and coupling modules, we can design arbitrary motions in large networks. If the modules preserve some symmetries in their motions, this coupling can be performed much more efficiently through the combining of nodes to create materials that replicate the module motion on a larger scale (see supplementary methods). We note that this procedure is simply extended to the design of networks with multiple motions $\bm{u}_{S1}, \bm{u}_{S2}$ by constraining the primary modules and the full network to have $d(d+1)/2 + 2$ degrees of freedom.

\section{Avoiding States of Self Stress in 3 Dimensions}
One crucial condition to guarantee finitely deformable motions is to avoid states of self stress during the judicious constraint process. A peculiar situation arises when designing networks with 5 specified node positions and motions $|\mc{V}_S| = 5$ in $d = 3$. In general, for 5 specified nodes with independent motions, we have $\dim(\tilde{A}) = 2$, with a one dimensional solution space that is the intersection of a quadric surface (defined by the first four nodes) and a plane (defined by the last node). Hence, all unspecified nodes in $\mc{V}_U$ must be placed coplanar to each other, which creates a bipartite network that provably has at least two infinitesimal motions. These motions mean that if we judiciously constrain our 5 specified nodes to try to achieve a total $N_D = 7$, we end up with 6 rigid body motions, 2 infinitesimal motions, and 1 state of self stress.

\begin{figure}[h!]
	\centering
	\includegraphics[width=1.0\columnwidth]{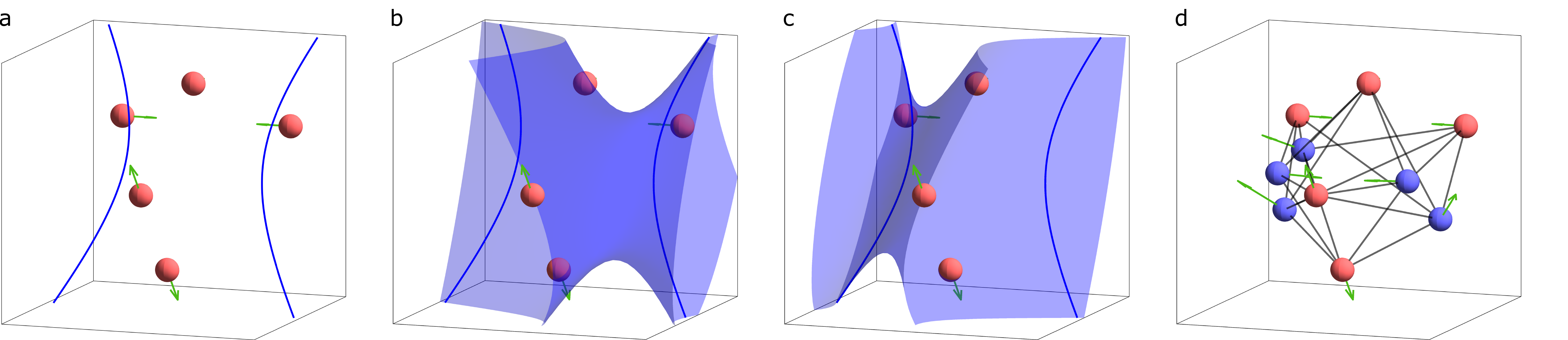}
	\caption{\textbf{Non-Planar Judicious Constraint of 5 Nodes in $d=3$.} (\textbf{a}) Planar 1 dimensional solution space (blue curves) for the position and motion of an unspecified node connected to 5 independent specified nodes. (\textbf{b}) 2 dimensional solution quadric surface (blue) for one subset of four of the five specified nodes, and (\textbf{c}) for a different subset. (\textbf{d}) Final constrained network with $N_D = 7$ degrees of freedom and no states of self stress, with the finitely deformable conformational motion in green arrows.}
	\label{fig:mn_combine_mod}
\end{figure}

For example, consider one of the modules in the main text in $d=3$ concerning network combination. The desired positions and motions of the specified nodes are
\begin{align*}
\bm{x}_1 = \begin{bmatrix} 2\\0\\1 \end{bmatrix},
\bm{x}_2 = \begin{bmatrix} 1\\0\\2 \end{bmatrix},
\bm{x}_3 = \begin{bmatrix} 3\\0\\3 \end{bmatrix},
\bm{x}_4 = \begin{bmatrix} 3.5\\-1\\2.5 \end{bmatrix},
\bm{x}_5 = \begin{bmatrix} 3.5\\1\\2.5 \end{bmatrix},
\end{align*}
\begin{align*}
\bm{u}_1 = \begin{bmatrix} -0.8\\0\\0.8 \end{bmatrix},
\bm{u}_2 = \begin{bmatrix} 0.8\\0\\-0.8 \end{bmatrix},
\bm{u}_3 = \begin{bmatrix} -0.8\\0\\0 \end{bmatrix},
\bm{u}_4 = \begin{bmatrix} 0\\-0.8\\0 \end{bmatrix},
\bm{u}_5 = \begin{bmatrix} 0\\0.8\\0 \end{bmatrix}.
\end{align*}
The motions were scaled to 0.8 for purely aesthetic reasons so that the figure arrows would not overlap. Here, we solve for and visually demonstrate that the 1 dimensional solution space lies along a plane (Fig.~\ref{fig:mn_combine_mod}a), such that even if we added $N_U = 4$, $N_B = 20$ to theoretically $N_D = 3(5+4)-20 = 7$, the generated state of self stress would not guarantee our desired motion as the sole finitely deformable motion. 

In response, we can add three coplanar unspecified nodes connected to all five specified nodes along the 1 dimensional solution space (Fig~\ref{fig:mn_combine_mod}a) to yield $N_D = 3(5+3) - 15 = 9$. Then we can remove the final two degrees of freedom by judiciously constraining two separate subsets of 4 specified nodes (Fig.~\ref{fig:mn_combine_mod}bc) along the quadric surface of solutions that are not coplanar to the initial 3 unspecified nodes, to get our final network $N_D = 7$, with the one desired finitely deformable conformation degree of freedom (Fig.~\ref{fig:mn_combine_mod}d).


\begin{thebibliography}{10}
	\expandafter\ifx\csname url\endcsname\relax
	\def\url#1{\texttt{#1}}\fi
	\expandafter\ifx\csname urlprefix\endcsname\relax\def\urlprefix{URL }\fi
	\providecommand{\bibinfo}[2]{#2}
	\providecommand{\eprint}[2][]{\url{#2}}
	
	\bibitem{Papadopoulos2017network}
	\bibinfo{author}{Papadopoulos, L.}, \bibinfo{author}{Porter, M.~A.},
	\bibinfo{author}{Daniels, K.~E.} \& \bibinfo{author}{Bassett, D.~S.}
	\newblock \bibinfo{title}{Network analysis of particles and grains}.
	\newblock \emph{\bibinfo{journal}{arXiv}} \textbf{\bibinfo{volume}{1708}},
	\bibinfo{pages}{08080} (\bibinfo{year}{2017}).
	
	\bibitem{Picu2011mechanics}
	\bibinfo{author}{Picu, R.~C.}
	\newblock \bibinfo{title}{Mechanics of random fiber networks - a review}.
	\newblock \emph{\bibinfo{journal}{Soft Matter}} \textbf{\bibinfo{volume}{7}},
	\bibinfo{pages}{6768--6785} (\bibinfo{year}{2011}).
	
	\bibitem{Vermeulen2017geometry}
	\bibinfo{author}{Vermeulen, M. F.~J.}, \bibinfo{author}{Bose, A.},
	\bibinfo{author}{Storm, C.} \& \bibinfo{author}{Ellenbroek, W.~G.}
	\newblock \bibinfo{title}{Geometry and the onset of rigidity in a disordered
		network}.
	\newblock \emph{\bibinfo{journal}{Phys. Rev. E}} \textbf{\bibinfo{volume}{96}},
	\bibinfo{pages}{053003} (\bibinfo{year}{2017}).
	
	\bibitem{bassett2012influence}
	\bibinfo{author}{Bassett, D.~S.}, \bibinfo{author}{Owens, E.~T.},
	\bibinfo{author}{Daniels, K.~E.} \& \bibinfo{author}{Porter, M.~A.}
	\newblock \bibinfo{title}{Influence of network topology on sound propagation in
		granular materials}.
	\newblock \emph{\bibinfo{journal}{Phys Rev E}} \textbf{\bibinfo{volume}{86}},
	\bibinfo{pages}{041306} (\bibinfo{year}{2012}).
	
	\bibitem{shi2014network}
	\bibinfo{author}{Shi, F.}, \bibinfo{author}{Wang, S.}, \bibinfo{author}{Forest,
		M.~G.} \& \bibinfo{author}{Mucha, P.~J.}
	\newblock \bibinfo{title}{Network-based assessments of percolation-induced
		current distributions in sheared rod macromolecular dispersions}.
	\newblock \emph{\bibinfo{journal}{Multiscale Modeling and Simulation}}
	\textbf{\bibinfo{volume}{12}}, \bibinfo{pages}{249--264}
	(\bibinfo{year}{2014}).
	
	\bibitem{norton2003design}
	\bibinfo{author}{Norton, R.}
	\newblock \emph{\bibinfo{title}{Design of Machinery: An Introduction to the
			Synthesis and Analysis of Mechanisms and Machines}}
	(\bibinfo{publisher}{McGraw-Hill Higher Education}, \bibinfo{year}{2003}).
	
	\bibitem{detweiler2007robotics}
	\bibinfo{author}{Detweiler, C.}, \bibinfo{author}{Vona, M.},
	\bibinfo{author}{Yoon, Y.}, \bibinfo{author}{{Seung-Kook Yun}} \&
	\bibinfo{author}{Rus, D.}
	\newblock \bibinfo{title}{{Self-assembling mobile linkages}}.
	\newblock \emph{\bibinfo{journal}{IEEE Robotics {\&} Automation Magazine}}
	\textbf{\bibinfo{volume}{14}}, \bibinfo{pages}{45--55}
	(\bibinfo{year}{2007}).
	
	\bibitem{patek2007mantis}
	\bibinfo{author}{Patek, S.~N.}, \bibinfo{author}{Nowroozi, B.~N.},
	\bibinfo{author}{Baio, J.~E.}, \bibinfo{author}{Caldwell, R.~L.} \&
	\bibinfo{author}{Summers, A.~P.}
	\newblock \bibinfo{title}{{Linkage mechanics and power amplification of the
			mantis shrimp's strike}}.
	\newblock \emph{\bibinfo{journal}{Journal of Experimental Biology}}
	\textbf{\bibinfo{volume}{210}}, \bibinfo{pages}{3677--3688}
	(\bibinfo{year}{2007}).
	
	\bibitem{crapo1979structural}
	\bibinfo{author}{Crapo, H.}
	\newblock \bibinfo{title}{Structural rigidity}.
	\newblock \emph{\bibinfo{journal}{Structural Topology}}
	\textbf{\bibinfo{volume}{73}}, \bibinfo{pages}{26--45}
	(\bibinfo{year}{1979}).
	
	\bibitem{Maxwell1864}
	\bibinfo{author}{Maxwell, J.~C.}
	\newblock \bibinfo{title}{{On the calculation of the equilibrium and stiffness
			of frames}}.
	\newblock \emph{\bibinfo{journal}{Philosophical Magazine Series 4}}
	\textbf{\bibinfo{volume}{27}}, \bibinfo{pages}{294--299}
	(\bibinfo{year}{1864}).
	
	\bibitem{Grimm1975}
	\bibinfo{author}{Grimm, H.} \& \bibinfo{author}{Dorner, B.}
	\newblock \bibinfo{title}{{On the mechanism of the $\alpha$-$\beta$ phase
			transformation of quartz}}.
	\newblock \emph{\bibinfo{journal}{Journal of Physics and Chemistry of Solids}}
	\textbf{\bibinfo{volume}{36}}, \bibinfo{pages}{407--413}
	(\bibinfo{year}{1975}).
	
	\bibitem{Hammonds1996}
	\bibinfo{author}{Hammonds, K.~D.}, \bibinfo{author}{Dove, M.~T.},
	\bibinfo{author}{Giddy, A.~P.}, \bibinfo{author}{Heine, V.} \&
	\bibinfo{author}{Winkler, B.}
	\newblock \bibinfo{title}{{Rigid-unit phonon modes and structural phase
			transitions in framework silicates}}.
	\newblock \emph{\bibinfo{journal}{American Mineralogist}}
	\textbf{\bibinfo{volume}{81}}, \bibinfo{pages}{1057--1079}
	(\bibinfo{year}{1996}).
	
	\bibitem{Broedersz2011}
	\bibinfo{author}{Broedersz, C.~P.}, \bibinfo{author}{Mao, X.},
	\bibinfo{author}{Lubensky, T.~C.} \& \bibinfo{author}{MacKintosh, F.~C.}
	\newblock \bibinfo{title}{{Criticality and isostaticity in fibre networks}}.
	\newblock \emph{\bibinfo{journal}{Nature Physics}}
	\textbf{\bibinfo{volume}{7}}, \bibinfo{pages}{983--988}
	(\bibinfo{year}{2011}).
	
	\bibitem{Sharma2016}
	\bibinfo{author}{Sharma, A.} \emph{et~al.}
	\newblock \bibinfo{title}{{Strain-controlled criticality governs the nonlinear
			mechanics of fibre networks}}.
	\newblock \emph{\bibinfo{journal}{Nature Physics}}
	\textbf{\bibinfo{volume}{12}}, \bibinfo{pages}{584--587}
	(\bibinfo{year}{2016}).
	
	\bibitem{lubensky2015phonons}
	\bibinfo{author}{Lubensky, T.~C.}, \bibinfo{author}{Kane, C.~L.},
	\bibinfo{author}{Mao, X.}, \bibinfo{author}{Souslov, A.} \&
	\bibinfo{author}{Sun, K.}
	\newblock \bibinfo{title}{Phonons and elasticity in critically coordinated
		lattices}.
	\newblock \emph{\bibinfo{journal}{Reports on Progress in Physics}}
	\textbf{\bibinfo{volume}{78}} (\bibinfo{year}{2015}).
	
	\bibitem{Bertoldi2017}
	\bibinfo{author}{Bertoldi, K.}, \bibinfo{author}{Vitelli, V.},
	\bibinfo{author}{Christensen, J.} \& \bibinfo{author}{van Hecke, M.}
	\newblock \bibinfo{title}{{Flexible mechanical metamaterials}}.
	\newblock \emph{\bibinfo{journal}{Nature Reviews Materials}}
	\textbf{\bibinfo{volume}{2}}, \bibinfo{pages}{17066} (\bibinfo{year}{2017}).
	
	\bibitem{Guo2016protein}
	\bibinfo{author}{Guo, J.} \& \bibinfo{author}{Zhou, H.~X.}
	\newblock \bibinfo{title}{Protein allostery and conformational dynamics}.
	\newblock \emph{\bibinfo{journal}{Chem Rev}} \textbf{\bibinfo{volume}{116}},
	\bibinfo{pages}{6503--6515} (\bibinfo{year}{2016}).
	
	\bibitem{kempe1875trajectory}
	\bibinfo{author}{Kempe, A.~B.}
	\newblock \bibinfo{title}{{On a General Method of describing Plane Curves of
			the n th degree by Linkwork}}.
	\newblock \emph{\bibinfo{journal}{Proceedings of the London Mathematical
			Society}} \textbf{\bibinfo{volume}{s1-7}}, \bibinfo{pages}{213--216}
	(\bibinfo{year}{1875}).
	
	\bibitem{hartenberg1964linkage}
	\bibinfo{author}{Hartenberg, R. S. R.~S.} \& \bibinfo{author}{Denavit, J.}
	\newblock \emph{\bibinfo{title}{Kinematic synthesis of linkages}}
	(\bibinfo{publisher}{New York : McGraw-Hill}, \bibinfo{year}{1964}).
	
	\bibitem{kempe1877line}
	\bibinfo{author}{Kempe, A.~B.}
	\newblock \bibinfo{title}{{How to draw a straight line}}.
	\newblock \emph{\bibinfo{journal}{Nature}} \textbf{\bibinfo{volume}{16}},
	\bibinfo{pages}{65----67, 86----89, 125----127, and 145----146}
	(\bibinfo{year}{1877}).
	
	\bibitem{Rocks2017}
	\bibinfo{author}{Rocks, J.~W.} \emph{et~al.}
	\newblock \bibinfo{title}{{Designing allostery-inspired response in mechanical
			networks}}.
	\newblock \emph{\bibinfo{journal}{Proceedings of the National Academy of
			Sciences}} \textbf{\bibinfo{volume}{114}}, \bibinfo{pages}{2520--2525}
	(\bibinfo{year}{2017}).
	
	\bibitem{goodrich2015pruning}
	\bibinfo{author}{Goodrich, C.~P.}, \bibinfo{author}{Liu, A.~J.} \&
	\bibinfo{author}{Nagel, S.~R.}
	\newblock \bibinfo{title}{{The Principle of Independent Bond-Level Response:
			Tuning by Pruning to Exploit Disorder for Global Behavior}}.
	\newblock \emph{\bibinfo{journal}{Physical Review Letters}}
	\textbf{\bibinfo{volume}{114}}, \bibinfo{pages}{225501}
	(\bibinfo{year}{2015}).
	\newblock \eprint{1502.02953}.
	
	\bibitem{Yan2017}
	\bibinfo{author}{Yan, L.}, \bibinfo{author}{Ravasio, R.},
	\bibinfo{author}{Brito, C.} \& \bibinfo{author}{Wyart, M.}
	\newblock \bibinfo{title}{{Architecture and coevolution of allosteric
			materials}}.
	\newblock \emph{\bibinfo{journal}{Proceedings of the National Academy of
			Sciences}} \textbf{\bibinfo{volume}{114}}, \bibinfo{pages}{2526--2531}
	(\bibinfo{year}{2017}).
	
	\bibitem{roth1981}
	\bibinfo{author}{Roth, B.}
	\newblock \bibinfo{title}{{Rigid and Flexible Frameworks}}.
	\newblock \emph{\bibinfo{journal}{The American Mathematical Monthly}}
	\textbf{\bibinfo{volume}{88}}, \bibinfo{pages}{6} (\bibinfo{year}{1981}).
	
	\bibitem{whiteley1984}
	\bibinfo{author}{Whiteley, W.}
	\newblock \bibinfo{title}{{Infinitesimal motions of a bipartite framework}}.
	\newblock \emph{\bibinfo{journal}{Pacific Journal of Mathematics}}
	\textbf{\bibinfo{volume}{110}}, \bibinfo{pages}{233--255}
	(\bibinfo{year}{1984}).
	
	\bibitem{bolker1980}
	\bibinfo{author}{Bolker, E.} \& \bibinfo{author}{Roth, B.}
	\newblock \bibinfo{title}{{When is a bipartite graph a rigid framework?}}
	\newblock \emph{\bibinfo{journal}{Pacific Journal of Mathematics}}
	\textbf{\bibinfo{volume}{90}}, \bibinfo{pages}{27--44}
	(\bibinfo{year}{1980}).
	
	\bibitem{guest2006stiffness}
	\bibinfo{author}{Guest, S.}
	\newblock \bibinfo{title}{{The stiffness of prestressed frameworks: A unifying
			approach}}.
	\newblock \emph{\bibinfo{journal}{International Journal of Solids and
			Structures}} \textbf{\bibinfo{volume}{43}}, \bibinfo{pages}{842--854}
	(\bibinfo{year}{2006}).
	
	\bibitem{asimow1978rigidity}
	\bibinfo{author}{Asimow, L.} \& \bibinfo{author}{Roth, B.}
	\newblock \bibinfo{title}{{The Rigidity of Graphs}}.
	\newblock \emph{\bibinfo{journal}{Transactions of the American Mathematical
			Society}} \textbf{\bibinfo{volume}{245}}, \bibinfo{pages}{279}
	(\bibinfo{year}{1978}).
	
	\bibitem{macaulay1902elimination}
	\bibinfo{author}{MacAulay, F.~S.}
	\newblock \bibinfo{title}{{Some Formulae in Elimination}}.
	\newblock \emph{\bibinfo{journal}{Proceedings of the London Mathematical
			Society}} \textbf{\bibinfo{volume}{s1-35}}, \bibinfo{pages}{3--27}
	(\bibinfo{year}{1902}).
	
	\bibitem{changeux1998}
	\bibinfo{author}{Changeux, J.-P.} \& \bibinfo{author}{Edelstein, S.~J.}
	\newblock \bibinfo{title}{{Allosteric Receptors after 30 Years}}.
	\newblock \emph{\bibinfo{journal}{Neuron}} \textbf{\bibinfo{volume}{21}},
	\bibinfo{pages}{959--980} (\bibinfo{year}{1998}).
	
	\bibitem{allewell1989}
	\bibinfo{author}{Allewell, N.~M.}
	\newblock \bibinfo{title}{{Escherichia Coli Aspartate Transcarbamoylase:
			Structure, Energetics, and Catalytic and Regulatory Mechanisms}}.
	\newblock \emph{\bibinfo{journal}{Annual Review of Biophysics and Biophysical
			Chemistry}} \textbf{\bibinfo{volume}{18}}, \bibinfo{pages}{71--92}
	(\bibinfo{year}{1989}).
	
	\bibitem{macol2001}
	\bibinfo{author}{Macol, C.~P.}, \bibinfo{author}{Tsuruta, H.},
	\bibinfo{author}{Stec, B.} \& \bibinfo{author}{Kantrowitz, E.~R.}
	\newblock \bibinfo{title}{{Direct structural evidence for a concerted
			allosteric transition in Escherichia coli aspartate transcarbamoylase.}}
	\newblock \emph{\bibinfo{journal}{Nature structural biology}}
	\textbf{\bibinfo{volume}{8}}, \bibinfo{pages}{423--6} (\bibinfo{year}{2001}).
	
	\bibitem{cockrell2013}
	\bibinfo{author}{Cockrell, G.~M.} \emph{et~al.}
	\newblock \bibinfo{title}{{New Paradigm for Allosteric Regulation of
			Escherichia coli Aspartate Transcarbamoylase}}.
	\newblock \emph{\bibinfo{journal}{Biochemistry}} \textbf{\bibinfo{volume}{52}},
	\bibinfo{pages}{8036--8047} (\bibinfo{year}{2013}).
	
	\bibitem{Miskin2013adapting}
	\bibinfo{author}{Miskin, M.~Z.} \& \bibinfo{author}{Jaeger, H.~M.}
	\newblock \bibinfo{title}{Adapting granular materials through artificial
		evolution}.
	\newblock \emph{\bibinfo{journal}{Nature Materials}}
	\textbf{\bibinfo{volume}{12}}, \bibinfo{pages}{326--331}
	(\bibinfo{year}{2013}).
	
	\bibitem{jacobs2001protein}
	\bibinfo{author}{Jacobs, D.~J.}, \bibinfo{author}{Rader, A.},
	\bibinfo{author}{Kuhn, L.~A.} \& \bibinfo{author}{Thorpe, M.}
	\newblock \bibinfo{title}{{Protein flexibility predictions using graph
			theory}}.
	\newblock \emph{\bibinfo{journal}{Proteins: Structure, Function, and Genetics}}
	\textbf{\bibinfo{volume}{44}}, \bibinfo{pages}{150--165}
	(\bibinfo{year}{2001}).
	
	\bibitem{paulose2015topological}
	\bibinfo{author}{Paulose, J.}, \bibinfo{author}{Chen, B. G.-g.} \&
	\bibinfo{author}{Vitelli, V.}
	\newblock \bibinfo{title}{{Topological modes bound to dislocations in
			mechanical metamaterials}}.
	\newblock \emph{\bibinfo{journal}{Nature Physics}}
	\textbf{\bibinfo{volume}{11}}, \bibinfo{pages}{153--156}
	(\bibinfo{year}{2015}).
	
	\bibitem{jacobs2012allostery}
	\bibinfo{author}{Jacobs, D.~J.} \emph{et~al.}
	\newblock \bibinfo{title}{{Ensemble Properties of Network Rigidity Reveal
			Allosteric Mechanisms}}.
	\newblock In \emph{\bibinfo{booktitle}{Methods in Molecular Biology}}, vol.
	\bibinfo{volume}{796}, \bibinfo{pages}{279--304} (\bibinfo{year}{2012}).
	
	\bibitem{silverberg2014origami}
	\bibinfo{author}{Silverberg, J.~L.} \emph{et~al.}
	\newblock \bibinfo{title}{{Using origami design principles to fold
			reprogrammable mechanical metamaterials}}.
	\newblock \emph{\bibinfo{journal}{Science}} \textbf{\bibinfo{volume}{345}},
	\bibinfo{pages}{647--650} (\bibinfo{year}{2014}).
	
	\bibitem{korner2015auxetic}
	\bibinfo{author}{K{\"{o}}rner, C.} \& \bibinfo{author}{Liebold-Ribeiro, Y.}
	\newblock \bibinfo{title}{{A systematic approach to identify cellular auxetic
			materials}}.
	\newblock \emph{\bibinfo{journal}{Smart Materials and Structures}}
	\textbf{\bibinfo{volume}{24}}, \bibinfo{pages}{025013}
	(\bibinfo{year}{2015}).
	
	\bibitem{lukin2004heme}
	\bibinfo{author}{Lukin, J.~A.} \& \bibinfo{author}{Ho, C.}
	\newblock \bibinfo{title}{{The Structure−Function Relationship of Hemoglobin
			in Solution at Atomic Resolution}}.
	\newblock \emph{\bibinfo{journal}{Chemical Reviews}}
	\textbf{\bibinfo{volume}{104}}, \bibinfo{pages}{1219--1230}
	(\bibinfo{year}{2004}).
	
	\bibitem{kamata2004allostery}
	\bibinfo{author}{Kamata, K.}, \bibinfo{author}{Mitsuya, M.},
	\bibinfo{author}{Nishimura, T.}, \bibinfo{author}{Eiki, J.-i.} \&
	\bibinfo{author}{Nagata, Y.}
	\newblock \bibinfo{title}{{Structural Basis for Allosteric Regulation of the
			Monomeric Allosteric Enzyme Human Glucokinase}}.
	\newblock \emph{\bibinfo{journal}{Structure}} \textbf{\bibinfo{volume}{12}},
	\bibinfo{pages}{429--438} (\bibinfo{year}{2004}).
	
	\bibitem{englander2014}
	\bibinfo{author}{Englander, S.~W.} \& \bibinfo{author}{Mayne, L.}
	\newblock \bibinfo{title}{{The nature of protein folding pathways}}.
	\newblock \emph{\bibinfo{journal}{Proceedings of the National Academy of
			Sciences}} \textbf{\bibinfo{volume}{111}}, \bibinfo{pages}{15873--15880}
	(\bibinfo{year}{2014}).
	
	\bibitem{papaleo2016}
	\bibinfo{author}{Papaleo, E.} \emph{et~al.}
	\newblock \bibinfo{title}{{The Role of Protein Loops and Linkers in
			Conformational Dynamics and Allostery}}.
	\newblock \emph{\bibinfo{journal}{Chemical Reviews}}
	\textbf{\bibinfo{volume}{116}}, \bibinfo{pages}{6391--6423}
	(\bibinfo{year}{2016}).
	
	\bibitem{lee2012auxetic}
	\bibinfo{author}{Lee, J.-H.}, \bibinfo{author}{Singer, J.~P.} \&
	\bibinfo{author}{Thomas, E.~L.}
	\newblock \bibinfo{title}{{Micro-/Nanostructured Mechanical Metamaterials}}.
	\newblock \emph{\bibinfo{journal}{Advanced Materials}}
	\textbf{\bibinfo{volume}{24}}, \bibinfo{pages}{4782--4810}
	(\bibinfo{year}{2012}).
	
	\bibitem{dagdelen2017search}
	\bibinfo{author}{Dagdelen, J.}, \bibinfo{author}{Montoya, J.},
	\bibinfo{author}{de~Jong, M.} \& \bibinfo{author}{Persson, K.}
	\newblock \bibinfo{title}{{Computational prediction of new auxetic materials}}.
	\newblock \emph{\bibinfo{journal}{Nature Communications}}
	\textbf{\bibinfo{volume}{8}}, \bibinfo{pages}{323} (\bibinfo{year}{2017}).
	
	\bibitem{yuzheng2005grasp}
	\bibinfo{author}{{Yu Zheng}} \& \bibinfo{author}{{Wen-Han Qian}}.
	\newblock \bibinfo{title}{{Dynamic force distribution in multifingered grasping
			by decomposition and positive combination}}.
	\newblock \emph{\bibinfo{journal}{IEEE Transactions on Robotics}}
	\textbf{\bibinfo{volume}{21}}, \bibinfo{pages}{718--726}
	(\bibinfo{year}{2005}).
	
\end{thebibliography}
\end{document}